\documentclass[structabstract]{aa}
\usepackage{graphicx}
\usepackage{txfonts}
\usepackage{natbib}
\bibpunct{(}{)}{;}{a}{}{,} 
\newcommand\curf{{\cal F}}
\begin{document}
\title{Accretion rates and accretion tracers of Herbig Ae/Be stars}

\author{I. Mendigut\'{\i}a \inst{1,2}
         \and
         N. Calvet\inst{3}
         \and
          B. Montesinos\inst{1}
         \and
         A. Mora\inst{4}
          \and
         J. Muzerolle\inst{5}
          \and
         C. Eiroa\inst{2}
         \and
         R.D. Oudmaijer\inst{6}
         \and
         B. Mer\'{\i}n\inst{7}}
         
\offprints{Ignacio Mendigut\'{\i}a\\
              \email{Ignacio.Mendigutia@cab.inta-csic.es}}
\institute{$^{1}$Centro de Astrobiolog\'{\i}a, Departamento de
     Astrof\'{\i}sica (CSIC-INTA), ESAC Campus, P.O. Box 78, 
     28691 Villanueva de la Ca\~nada, Madrid, Spain.\\
     $^{2}$Departamento de F\'{\i}sica Te\'{o}rica, M\'{o}dulo 15,
     Facultad de Ciencias, Universidad Aut\'{o}noma de Madrid, PO Box
     28049, Cantoblanco, Madrid, Spain.\\
     $^{3}$Department of Astronomy, University of Michigan, 830 Dennison
     Building, 500 Church Street, Ann Arbor, MI 48109\\
     $^{4}$GAIA Science Operations Centre, ESA, European Space Astronomy Centre, PO Box 78, 28691, Villanueva de la Ca\~nada, Madrid,
     Spain.\\
     $^{5}$Space Telescope Science Institute, 3700 San Martin Dr., Baltimore,
     MD, 21218\\
     $^{6}$School of Physics \& Astronomy, University of Leeds, Woodhouse
     Lane, Leeds LS2 9JT, UK.\\
     $^{7}$Herschel Science Centre, ESA, European Space Astronomy Centre, P.O. Box 78, 28691, Villanueva de la Ca\~nada, Madrid,
     Spain.\\}

     \date{Received 2011, June 9; accepted 2011, September 2}
 
  \abstract{The scarcity of accretion rate estimates and accretion tracers available for Herbig Ae/Be (HAeBe) stars contrasts with the extensive studies for lower mass objects.}
{This work aims to derive accretion rates from the UV Balmer excess for a sample of 38 HAeBe stars. We look for possible empirical correlations with the strength of the H$\alpha$, [\ion{O}{i}]6300, and Br$\gamma$ emission lines.}
{Shock modelling within the context of magnetospheric accretion (MA) was applied to each star. We obtained the accretion rates from the excess in the Balmer discontinuity, derived from mean values of multi-epoch Johnson's $UB$ photometry. The accretion rates were related to both mean H$\alpha$ luminosities, H$\alpha$ 10$\%$ widths, and [\ion{O}{i}]6300 luminosities from simultaneous spectra, and to Br$\gamma$ luminosities from the literature.}
{The typical -median- mass accretion rate is 2 $\times$ 10$^{-7}$ M$_{\sun}$ yr$^{-1}$ in our sample, 36 $\%$ of the stars showing values $\leq$ 10$^{-7}$ M$_{\sun}$ yr$^{-1}$, 35$\%$ between 10$^{-7}$ and 10$^{-6}$, and 29$\%$ $>$ 10$^{-6}$ M$_{\sun}$ yr$^{-1}$. The model fails to reproduce the large Balmer excesses shown by the four hottest stars (T$_{*}$ $>$ 12000 $K$). When accretion is related to the stellar masses and luminosities (1 $\leq$ M$_{*}$/M$_{\odot}$ $\leq$ 6; 2 $\leq$ L$_{*}$/L$_{\odot}$ $\leq$ 10$^3$), we derive $\dot{M}_{\rm acc}$ $\propto$ M$_{*}^{5}$ and $L$$_{\rm acc}$ $\propto$ L$_{*}^{1.2}$, with scatter. Empirical calibrations relating the accretion and the H$\alpha$, [\ion{O}{i}]6300, and Br$\gamma$ luminosities are provided. The slopes in our expressions are slightly shallower than those for lower mass stars, but the difference is within the uncertainties, except for the [\ion{O}{i}]6300 line. The H$\alpha$ 10$\%$ width is uncorrelated with $\dot{M}_{\rm acc}$, unlike for the lower mass regime. The mean H$\alpha$ width shows higher values as the projected rotational velocities of HAe stars increase, which agrees with MA. The accretion rate variations in the sample are typically lower than 0.5 dex on timescales of days to months, Our data suggest that the changes in the Balmer excess are uncorrelated to the simultaneous changes of the line luminosities.}
{The Balmer excesses and H$\alpha$ line widths of HAe stars can be interpreted within the context of MA, which is not the case for several HBes. The steep trend relating $\dot{M}_{\rm acc}$ and M$_{*}$ can be explained from the mass-age distribution characterizing HAeBe stars. The line luminosities used for low-mass objects are also valid to estimate typical accretion rates for the intermediate-mass regime under similar empirical expressions. However, we suggest that several of these calibrations are driven by the stellar luminosity.}

\maketitle
%
\section{Introduction}
\label{Section:Introduction}

Most of the stellar mass acquired during the pre-main sequence (PMS) phase comes through accretion from the disk. In addition, estimates of the mass accretion rate are needed to analyse circumstellar gas dissipation, and therefore to quantify the timescale when planets could be formed in protoplanetary disks. Magnetospheric accretion \citep[MA hereafter,][]{Uchida85,Konigl91,Shu94} is the accepted paradigm that explains disk-to-star accretion in classical T-Tauri (CTT) and lower mass stars. In the MA scenario, the inner disk is truncated at some point between the stellar surface and the co-rotation radius, where matter is accelerated through the magnetic field lines until it reaches the central star. The resulting hot accretion shocks can be modelled, allowing us to explain the continuum excess and the spectroscopic veiling from temperature differences with the stellar photosphere \citep{CalvetGullbrig98}. This modelling yields accretion rate estimates, which are found to correlate with the strength of spectroscopic features covering the wavelength range from the UV to the IR \citep[see e.g.][and references therein]{Herczeg08,Fang09,Rigliaco11}. Even though the origin of these correlations is not clear, those spectral lines are used as empirical tracers, which simplifies the process of estimating accretion rates. 

Herbig Ae/Be (HAeBe) stars are the massive (1-10M$_{\sun}$) counterparts of CTTs. How circumstellar matter accretes on those objects remains an open issue. For instance, MS stars with spectral types earlier than $\sim$ A6 (M$_{*}$ $\gtrsim$ 2 M$_{\sun}$) are not expected to have convective sub-photospheric zones generating the necessary magnetic fields \citep[][and references therein]{Simon02}. However, in young stars, convection zones may appear in earlier spectral types than expected \citep{Finkenzeller85}. In addition, different evidence suggests that MA could be extended to the intermediate-mass regime. Although weak in several cases, magnetic fields have been detected in some HAeBes \citep{Wade07,Hubrig09}. Spectropolarimetric measurements point to MA acting in HAe stars \citep{Vink02,Vink03,Mottram07}, and \citet{Grady10} conclude that the accretion in these objects goes through high-latitude funnels, again suggesting magnetically controlled accretion \citep[see also][]{Brit09}. High-velocity redshifted self-absorptions observed in line profiles of several HAeBe stars \citep{Natta00,Mora02,Mora04} point to infalling material at close to free-fall velocities. These can be easily explained from MA and hardly at all from a competing scenario such as the boundary-layer one \citep[see e.g. the review in][]{Alencar07}.  

MA modelling was carried out for a sample of nine intermediate-mass T-Tauri stars by \citet{Calvet04}, showing that the correlation between the accretion and Br$\gamma$ luminosities can be extended to these objects. The problem of measuring accretion rates for the HAeBe regime was faced by \citet[][MDCH04 hereafter]{Muzerolle04}. They reproduced the line profiles of \object{UX Ori} by assuming the MA geometry. The absence of significant line veiling and optical excess in most HAeBe objects was explained from the similar temperatures characterizing the stellar photosphere and the accretion shocks. The excess in the Balmer discontinuity was shown to be a valid measurement of accretion by MDCH04, providing a calibration between both parameters for a typical HAe star.

The large number of accretion studies for wide samples of low-mass stars contrasts with the lack of accretion rate measurements for HAeBes, which is also reflected in the scarce number of empirical tracers of accretion available for these objects. Accretion rate estimates for a wide sample of HAeBe stars were obtained by \citet{GarciaLopez06} from the Br$\gamma$ line, using the correlation derived by \citep{Calvet04} for stars in a lower range of stellar temperatures. The recent accretion rate estimates by \citet{DonBrit11} are based on the calibration between the mass accretion rate and the excess in the Balmer discontinuity provided in MDCH04, which was modelled for a specific set of stellar parameters. However, the measurement of accretion requires modelling observed properties for each object individually. The accretion rates can then be compared with spectral diagnostics to establish possible empirical correlations. That comparison could benefit from simultaneous measurements, given the variability characterizing PMS stars \citep[][Paper I hereafter]{Eiroa02,Mendi11}. 

Following MDCH04, in this work we apply shock modelling within the context of MA, aiming to reproduce the strength of mean Balmer excesses derived from multi-epoch photometry of a wide sample of HAeBe stars. The accretion rates obtained in this way are then related to the mean strength of the H$\alpha$ and [\ion{O}{i}]6300 emission lines from simultaneous spectra, as well as to non-simultaneous Br$\gamma$ luminosities from the literature. In addition, the simultaneous multi-epoch data allows us to make a first approach on the accretion rate variability and its relation with the spectral lines.  

Section \ref{Section:Sample} summarizes the properties of the sample and the data. Section \ref{section: accretion} describes the shock model (Sect. \ref{section:descriptmodel}) and provides the accretion rates estimates (Sect. \ref{section:discuss}), which are compared to previous results (Sect. \ref{section:prevresults}). Section \ref{section:actracers} relates our accretion luminosities with those of the H$\alpha$, [\ion{O}{i}]6300, and Br$\gamma$ lines. The relation between accretion and the H$\alpha$ 10$\%$ width is treated separately in Sect. \ref{section:W10}. The analysis of the accretion rate variability is outlined in Sect. \ref{section:accretionvar}. The discussion and conclusions are included in Sects. \ref{Section:Discussion} and \ref{Section:Conclusions}.

\section{Sample properties and data}
\label{Section:Sample}

Table \ref{Table:sample} shows stellar parameters of the 38 stars in the sample. Columns 1 to 9 indicate the name of the star, the stellar mass, luminosity, effective temperature, radius, surface gravity, age, projected rotational velocity ($v\sin i$), and distance. Uncertainties can be found in some of the references given in the caption. \citet{Montesinos08}, from where most of the values are taken, gives around 12$\%$, 40$\%$, $\pm$ 150 K, $\pm$ 0.1 dex and 35$\%$ and  for the stellar mass, luminosity, effective temperature, surface gravity and age, respectively. The low errors, as compared to other methods based on spectral types, arise from a detailed spectroscopic and photometric modelling, based on the comparison to Kurucz synthetic spectra and stellar evolutionary tracks. The typical uncertainty for $v\sin i$ is 6$\%$ \citep{Mora01}. The sample is the same one as was analysed in Paper I. The objects are mainly HAe and late-type HBes, as well as ten intermediate-mass T-Tauri stars with F--G spectral types and 1-4 M$_{\sun}$. All the stars show IR excess \citep{Merin04} indicative of dusty circumstellar environments, and the H$\alpha$ line in emission, pointing to active accretion.

This work is mainly based on the multi-epoch H$\alpha$ and [\ion{O}{i}]6300 spectra from Paper I, and the multi-epoch $UBV$ photometry from \citet{Oudmaijer01}. The spectra were taken at mid-resolution (R $\sim$ 5500), and the photospheric contribution was subtracted. We selected those spectra and photometry taken on the same nights. Therefore, the results in this paper were derived from (almost) simultaneous spectroscopic and photometric measurements, unless otherwise stated. The slit width of the spectrograph was 1'' projected in the sky, which avoids contamination from stellar companions in almost all sources. This is not the case for the photometry, where the aperture was $\sim$ 14''. However, most of the stars in our sample with reported multiplicity have much fainter companions in the optical \citep[see e.g. the references in][]{Wheelwright10}. We assume that their contribution to the emission at short wavelengths is negligible for our purposes. 

\begin{table*}
\centering
\renewcommand\tabcolsep{2.4pt}
\caption{Sample of stars.}
\label{Table:sample}
\begin{tabular}{llllllllllllll}
\hline\hline
Star & M$_{*}$ & L$_{*}$ & T$_{*}$ & R$_{*}$ & g & Age & $v\sin i$ & d & $<$L(H$\alpha$)$>$ & $<$W$_{10}$(H$\alpha$)$>$ & $<$L([OI]6300)$>$ & $<$E(B-V)$>$ & L(Br$\gamma$)\\ 
 & (M$_{\sun}$) & (L$_{\sun}$) & (K) & (R$_{\sun}$) & [cm s$^{-2}$] & (Myr)&(km s$^{-1}$) & (pc) &[L$_{\sun}$] & (km s$^{-1}$) & [L$_{\sun}$] &(mag) & [L$_{\sun}$]\\
\hline
HD 31648  & 2.0          & 21.9       & 8250        &2.3&4.0&6.7	   &102      &146 &-1.42&595&...  &0.02& (?) \\
HD 34282  & $<$2.1$^{A}$ & 5.13$^{A}$ & 9550$^{A}$  &0.8&4.9&$>$ 7.8$^{A}$ &129      &164$^A$ &-2.82&487&...  &0.19&-4.20$^{1}$\\
HD 34700  & 2.4$^{B}$    & 20.0$^{B}$ & 6000$^{B}$  &4.2&3.6&3.4$^{B}$     &46       &336$^H$ &-2.29&334&...  &0.01&(?)  \\
HD 58647  & 6.0          & 911        & 10500       &9.1&3.3&0.4	   &118      &543 &-0.13&619&-2.49&0.13&-2.08$^{2}$\\
HD 141569 & 2.2$^{A}$    & 22.9$^{A}$ & 9550$^{A}$  &1.8&4.3&6.7$^{A}$     &258      &99$^A$  &-2.01&646&-3.71&0.09&-3.99$^{1,2}$\\
HD 142666 & 2.0$^{A}$    & 17.0$^{A}$ & 7590$^{A}$  &2.4&4.0&5.1$^{A}$     &72       &145$^A$ &-2.33&483&-4.75&0.26&-3.53$^{1,3}$\\
HD 144432 & 2.0$^{A}$    & 14.8$^{A}$ & 7410$^{A}$  &2.3&4.0&5.3$^{A}$     &85       &145$^A$ &-1.87&421&-4.93&0.06&-3.29$^{1,3}$\\
HD 150193 & 2.2          & 36.1       & 8970        &2.5&4.0&5.0	   &100$^{C}$&203 &-1.15&458&...  &0.45&-2.64$^{1}$\\
HD 163296 & 2.2          & 34.5       & 9250        &2.3&4.1&5.0	   &133      &130 &-1.17&726&-4.37&0.03&-2.77$^{1,2,3}$\\
HD 179218 & 2.6          & 63.1       & 9500        &2.9&3.9&3.3	   &72$^{D}$ &201 &-1.16&464&-3.86&0.08&-2.74$^{3}$\\
HD 190073 & 5.1          & 471        & 9500        &8.0&3.4&0.6	   &20$^{E}$ &767 &0.06 &378&-2.49&0.13&(?)  \\
AS 442    & 3.5          & 207        & 11000       &4.0&3.8&1.5	   &(?)      &826 &-0.15&646&-2.42&0.73&(?)  \\
VX Cas    & 2.3          & 30.8       & 10000       &1.9&4.3&6.4	   &179      &619 &-1.43&672&-3.48&0.37&(?)  \\
BH Cep    & 1.7$^{A}$    & 8.91$^{A}$ & 6460$^{A}$  &2.4&3.9&8.2$^{A}$     &98       &450$^A$ &-2.34&705&-4.25&0.31&(?)  \\
BO Cep    & 1.5$^{A}$    & 6.61$^{A}$ & 6610$^{A}$  &2.0&4.0&11.2$^{A}$    &(?)      &400$^A$ &-2.51&685&-3.97&0.13&(?)  \\
SV Cep    & 2.4          & 37.5       & 10250       &1.9&4.3&5.2	   &206      &596 &-1.33&731&-3.20&0.39&(?)  \\
V1686 Cyg & $>$3.5$^{A}$ & 257$^{A}$  & 6170$^{A}$  &14 &2.7&$<$ 0.2$^{A}$ &(?)      &980$^A$ &-0.27&457&-2.80&0.63&-1.77$^{3}$\\
R Mon     & $>$5.1$^{A}$ & 2690$^{A}$ & 12020$^{A}$ &12 &3.0&$<$ 0.01$^{A}$&(?)      &800$^A$ &0.34 &832&-1.04&0.70&(?)  \\
VY Mon    & $>$5.1$^{A}$ & 15800$^{A}$& 12020$^{A}$ &29 &2.5&$<$ 0.01$^{A}$&(?)      &800$^A$ &-0.65&719&-0.46&1.79&(?)  \\
51 Oph    & 4.2          & 312        & 10250       &5.6&3.6&0.7	   &256      &142 &-1.23&522&...  &0.03&-2.68$^{1,2}$\\
KK Oph    & 2.2$^{A}$    & 25.7$^{A}$ & 7590$^{A}$  &2.9&3.8&3.9$^{A}$     &177      &160$^A$ &-2.28&593&-3.53&0.36&-3.53$^{1}$\\
T Ori     & 2.4          & 50.2       & 9750        &2.5&4.0&4.0	   &175      &472 &-0.88&680&-2.95&0.54&(?)  \\
BF Ori    & 2.6          & 61.6       & 8970        &3.3&3.8&3.2	   &37       &603 &-1.24&731&-3.49&0.15&-2.92$^{3}$\\
CO Ori    & $>$3.6$^{A}$ & 100$^{A}$  & 6310$^{A}$  &8.4&3.1&$<$ 0.1$^{A}$ &65       &450$^A$ &-0.99&553&-2.77&0.70&(?)  \\
HK Ori    & 3.0$^{A}$    & 77.6$^{A}$ & 8510$^{A}$  &4.1&3.7&1.0$^{A}$     &(?)      &460$^A$ &-1.57&573&-2.69&0.37&-2.92$^{1,3}$\\
NV Ori    & 2.2$^{F}$    & 21.2$^{F}$ & 6750$^{F}$  &3.4&3.7&4.4$^{F}$     &81       &450$^I$ &-1.97&583&-4.81&0.08&(?)  \\
RY Ori    & 2.5$^{A}$    & 28.2$^{A}$ & 6310$^{A}$  &4.5&3.5&1.8$^{A}$     &66       &460 &-1.7 &598&-3.74&0.49&(?)  \\
UX Ori    & 2.3          & 36.8       & 8460        &2.8&3.9&4.5	   &215      &517 &-1.36&677&-3.58&0.17&-2.80$^{1,3}$\\
V346 Ori  & 2.5          & 61.4       & 9750        &2.8&4.0&3.5	   &(?)      &586 &-1.87&889&...  &0.29&-3.21$^{3}$\\
V350 Ori  & 2.2          & 29.3       & 8970        &2.2&4.1&5.5	   &(?)      &735 &-1.39&724&-3.26&0.47&-2.62$^{3}$\\
XY Per    & 2.8          & 85.6       & 9750        &3.3&3.9&2.5	   &217      &347 &-1.12&728&-3.29&0.46&-2.97$^{3}$\\
VV Ser    & 4.0          & 336        & 13800       &3.2&4.0&1.2	   &229      &614 &-0.06&691&-1.82&1.04&-1.34$^{1,3}$\\
CQ Tau    & 1.5$^{B}$    & 5$^{B}$    & 6800$^{B}$  &1.6&4.2&7.7$^{B}$     &105      &130$^1$ &-2.86&529&-4.54&0.25&-3.96$^{1,3}$\\
RR Tau    & 5.8          & 781        & 10000       &9.3&3.3&0.4	   &225      &2103&0.04 &681&-1.47&0.51&-1.58$^{1}$\\
RY Tau    & 1.3$^{F}$    & 2.30$^{F}$ & 5770$^{F}$  &1.5&4.2&6.5$^{G}$     &55       &134$^I$ &-1.97&648&-3.43&0.37&(?)  \\
PX Vul    & 1.5$^{A}$    & 5.25$^{A}$ & 6760$^{A}$  &1.7&4.2&14$^{A}$	   &(?)      &420$^A$ &-1.56&628&-3.71&0.45&-2.80$^{3}$\\
WW Vul    & 2.5          & 50.0       & 8970        &2.9&3.9&3.7	   &220      &696 &-0.96&754&-3.17&0.36&-2.33$^{1}$\\
LkHa 234  & $>$5.3$^{A}$ & 4680$^{A}$ & 12900$^{A}$ &14 &2.9&$<$ 0.01$^{A}$&(?)      &1250$^A$&0.58 &747&-1.60&1.02&(?)  \\

\hline
\end{tabular}

\begin{minipage}{18cm}

  \textbf{Notes.} Surface gravities and line luminosities are on log scale. ``...'' means non-detections and ``(?)'' unknown values. Unless otherwise stated, the stellar masses, luminosities, effective temperatures, gravities, ages, and distances are from \citet{Montesinos08}; $v\sin i$ values from \citet{Mora01}, H$\alpha$ and [\ion{O}{i}]6300 mean values, and mean colour excesses from this work. Stellar radius are computed from M$_{*}$ and log g when they are available in \citet{Montesinos08}, obtaining equal values if derived from L$_{*}$ and T$_{*}$. R$_{*}$ comes from these parameters for the remaining stars, for which log g is then obtained from M$_{*}$ and R$_{*}$. The photometry for \object{AS 442} and R Mon was taken from SIMBAD (http://simbad.u-strasbg.fr/simbad/). Br$\gamma$ luminosities are from $^{1}$\citet{GarciaLopez06}, $^{2}$\citet{Brit07} and $^{3}$\citet{DonBrit11}. $^A$\citet{Manoj06}, $^B$\citet{AlonsoAlbi09}, $^C$\citet{Glebocki00}, $^D$\citet{Guimaraes06}, $^E$\citet{Hoffleit82}, $^F$\citet{Merin04}, $^G$\citet{Siess99}, $^H$\citet{Acke05}, $^I$\citet{Blondel06}.   
\end{minipage}
\end{table*} 

Columns 10, 11, and 12 of Table \ref{Table:sample} list the dereddened mean H$\alpha$ luminosities ($<$L(H$\alpha$)$>$), the mean H$\alpha$ widths at 10$\%$ of peak intensity ($<$W$_{10}$(H$\alpha$)$>$), and the dereddened mean [\ion{O}{i}]6300 luminosities ($<$L([\ion{O}{i}]6300)$>$). The mean line luminosities from Paper I were dereddened using the mean colour excesses in Col. 13, a total-to-selective extinction ratio R$_{\rm V}$ = 5 \citep{Hernandez04}, and the extinction law compiled in \citet{Robitaille07}\footnote{The extinction law can be downloaded from the web-based SED fitting tool in http://caravan.astro.wisc.edu/protostars/} from \citet{Kim94} and \citet{Indebetouw05}. The mean colour excesses were computed using the $BV$ photometry in \citet{Oudmaijer01} and the corresponding intrinsic colours in \citet{KenyonHartmann95}. Using a different colour to characterize the reddening in the sample \citep[e.g. $E(V-R)$, see][]{Calvet04} would not significantly affect our results. $<$E(B-V)$>$ is consistent with the colour excess applied by \citet{Montesinos08} to derive most of the stellar parameters used in this work. Regarding the total-to-selective extinction ratio, \citet{Hernandez04} confirmed that R$_{\rm V}$ = 5 is more consistent with the observed properties of HAeBe stars than the typical R$_{\rm V}$ = 3.1 for the interstellar medium. Using a different R$_{\rm V}$ or extinction law \citep[see e.g. the discussion in][]{Calvet04} would not change the major conclusions of this work, given that most objects in the sample are not heavily extincted (A$_{\rm V}$ $\leq$ 2.5 magnitudes for 79$\%$ of the stars). It is noted that different values for R$_{\rm V}$ were used to derive the stellar parameters in Table \ref{Table:sample} -e.g. R$_{\rm V}$ = 3.1 in \citet{Montesinos08} and a range 1 $\leq$ R$_{\rm V}$ $\leq$ 11 in \citet{Manoj06}-. Section \ref{section:actracers} again considers the influence of the R$_{\rm V}$ value on our results.

Finally. non-simultaneous Br$\gamma$ luminosities from the literature (see notes to Table \ref{Table:sample}) are given in Col. 14, with mean values adopted for the stars where two or three measurements are available. For an appropriate comparison, the Br$\gamma$ luminosities were rescaled to the same distances used in Paper I to derive the remaining line luminosities (Col. 9).

\section{Accretion rates}
\label{section: accretion}

\subsection{Description of the model}
\label{section:descriptmodel}
HAeBe stars show excess of continuum emission compared with MS stars with similar spectral types, in particular, at the Balmer discontinuity region \citep[see below and][]{Garrison78,DonBrit11}. The Balmer excesses are modelled in this section to provide estimates of the accretion rates. The analysis of the origin and strength of the magnetic fields necessary for driving accretion is beyond the scope of this work. We assume the MA geometry, using shock models similar to those successfully applied for the lower-mass regime. Following \citet{CalvetGullbrig98} and MDCH04, the total flux per wavelength unit emerging from the star is
\begin{equation}
\label{Eq:totflux}
F_{\lambda} = fF_{\lambda}^{col} + (1 - f)F_{\lambda}^{phot}
\end{equation}
where $f$ is the filling factor that reflects the stellar surface coverage of the accretion columns, $F$$_{\lambda}^{phot}$ the flux from the undisturbed photosphere, and $F$$_{\lambda}^{col}$ the flux from the column. The Kurucz model corresponding to a given T$_{*}$ and log g \citep{Kurucz93} is used to represent $F$$_{\lambda}^{phot}$\footnote{The use of a family of synthetic models different from Kurucz's does not significantly affect our results. The NextGen/PHOENIX models \citep{Hauschildt99} provide equal Balmer excesses, within 0.03 magnitudes, for the typical stellar and model parameters used (see text).}. In turn, the total luminosity of the columns is defined from the inward flux of energy carried by the accretion columns ($\curf$) and the outward stellar radiation below the accretion shocks \citep{CalvetGullbrig98}:
\begin{equation}
\label{Eq:condition1}
L^{col} = F^{col}A = (\curf + F^{phot})\times A = \xi L_{\rm acc} + F^{phot}A 
\end{equation}
where $A$ = $f$4$\pi$$\rm R$$_*^2$ is the surface area covered by the shocks, $L$$_{\rm acc}$ = GM$_{*}$$\dot{M}_{\rm acc}$/R$_*$ is the accretion luminosity; and $\xi$ = 1 - R$_*$/$R$$_i$, with $R$$_i$ the disk truncation radius. The parameter $F$$_{\lambda}^{col}$ is modelled as the flux from a blackbody at temperature $T$$_{col}$ (i.e. $F$$^{col}$ = $\sigma$$T$$_{col}^4$). This is justified for the range of stellar temperatures represented here, given that the main contribution to the flux in the optical and near-UV comes from the optically thick, heated photosphere and that the optically thin pre-shock contribution represents less than 1/4 of $L$$^{col}$ (see MDCH04). For the model parameters used in this work (see below), the blackbody approach improves as T$_{*}$ increases, and including the optically-thin contribution would change our modelled Balmer excesses by less than $\sim$ 0.07 magnitudes for the colder stars, which is accurate enough for our data. Since the total stellar luminosity can also be represented as a blackbody at the stellar temperature, $F$$^{phot}$ = $\sigma$T$_{*}^4$,  the second and third terms of Eq. \ref{Eq:condition1} provide

\begin{equation}
\label{Eq:F}
\sigma \it T_{col}^4 = \curf + \sigma \rm T_{*}^4. 
\end{equation} 
Finally, from the third and fourth terms of Eq. \ref{Eq:condition1}:
\begin{equation}
\label{Eq:condition2}
\curf \times f 4 \pi \rm R_*^2 = \left(1-\frac{\rm R_*}{\it R_i}\right)\frac{\rm {GM_{*}}\it \dot{M}_{\rm acc}}{\rm R_*}.
\end{equation}

Once $\curf$ and $R$$_i$ are fixed, the blackbody temperature characterizing the flux from the column and the filling factor for a given set of stellar and accreting parameters can be determined from Eqs. \ref{Eq:F} and \ref{Eq:condition2}, respectively. We fixed $\curf$ = 10$^{12}$ erg cm$^{-2}$ s$^{-1}$, which mostly provides appropriate filling factors \citep[$f$ $\lesssim$ 0.1; see e.g.][]{Valenti93} and reproduces most of the observed Balmer excesses (see Sect. \ref{section:discuss}). The use of a different value within the expected interval, 10$^{10}$ $\leq$ $\curf$(erg cm$^{-2}$ s$^{-1}$) $\leq$ 10$^{12}$, does not significantly affect the results (see MDCH04). The disk truncation radius should be smaller than the co-rotation one \citep[$R$$_{cor}$, see Sect. \ref{section:W10} and][]{Shu94}, which decreases with the stellar rotational velocity:

\begin{equation}
\label{Eq:Rcor}
R_{cor} = \left(\frac{\rm G M_* R_*^2}{v_*^2}\right)^{1/3}.
\end{equation}

The $v\sin i$ values from Table \ref{Table:sample} are used as proxies for $v$$_{*}$. The disk truncation radius was fixed to 2.5R$_{*}$, which is the expected value for HAeBe stars (MDCH04), or to $R$$_{cor}$ if this is smaller than 2.5R$_*$. The results do not significantly differ if R$_{*}$/$R$$_{i}$ is changed up to a factor 2 \citep{Herczeg08}.

\begin{figure*}
\centering
 \includegraphics[width=180mm,clip=true]{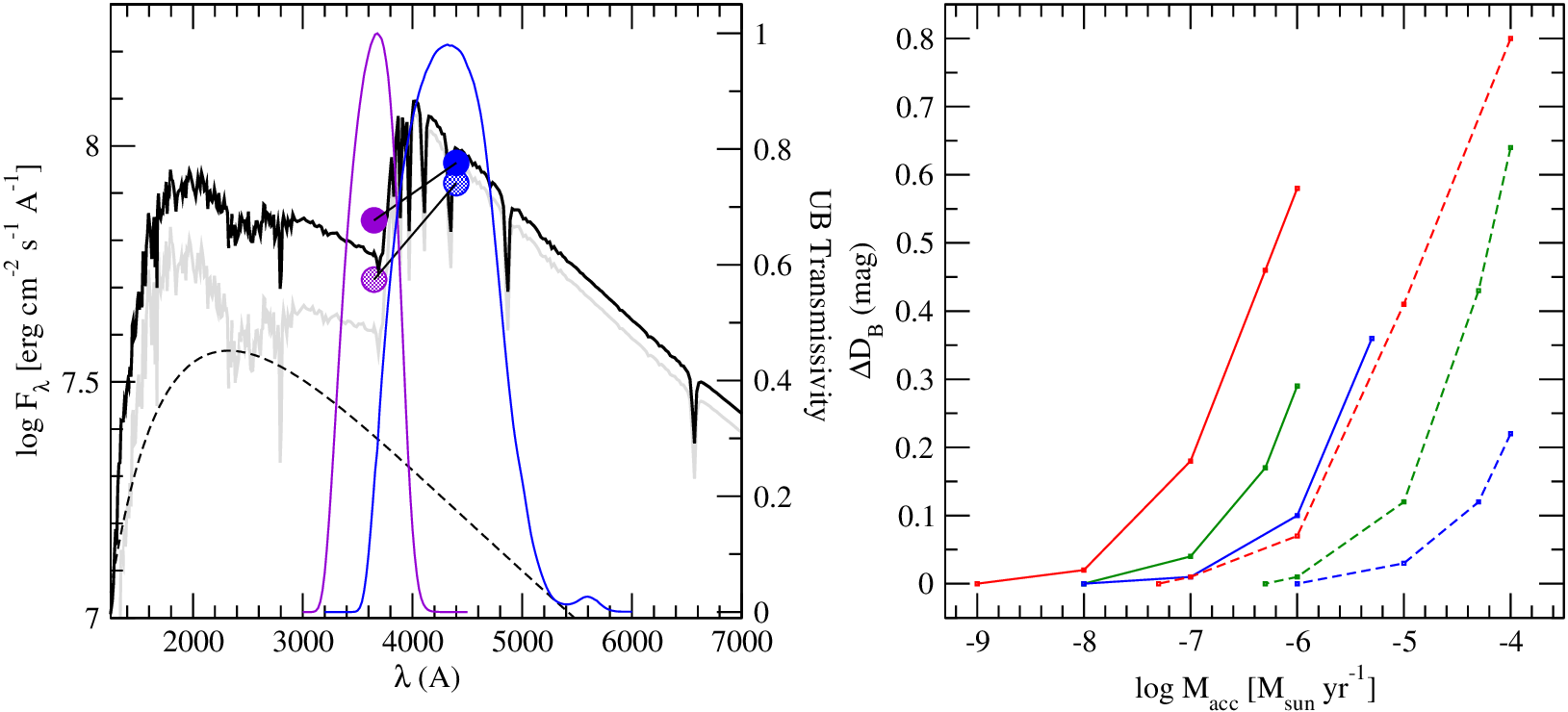}
\caption{(Left): Photospheric flux (grey line), contribution from accretion ($f$$F$$_{\lambda}^{col}$, dashed line) and total flux (solid black line) for M$_{*}$ = 2.5 M$_{\sun}$, R$_{*}$ = 2.6 R$_{\sun}$, T$_{*}$ = 9000 K, $\curf$ = 10$^{12}$ erg cm$^{-2}$ s$^{-1}$ ($T$$_{col}$ = 12470 K) and $R$$_i$ = 2.5R$_*$. The mass accretion rate is 5 $\times$ 10$^{-7}$ M$_{\sun}$ yr$^{-1}$,  ($f$ = 0.084). All fluxes are computed at the stellar surface. The transmission curves of the $U$ and $B$ filters are plotted in violet and blue, respectively. The convolution of the fluxes with the filter responses provide the synthetic photometric points used to obtain the modelled photospheric and total colours (from the shaded and filled linked circles, respectively). (Right): Predicted excess in the Balmer discontinuity as a function of mass accretion rate for T$_{*}$ = 6500, 9000, and 12500 K (red, green, and blue lines) and log g = 4.0 and 3.0 (solid and dashed lines). $\curf$ = 10$^{12}$ erg cm$^{-2}$ s$^{-1}$ and $R$$_i$ = 2.5R$_*$ for all cases.}
\label{Figure:balmeraccretion}
\end{figure*}

Once $T$$_{col}$ and $f$ are determined, the expected total flux can finally be obtained from Eq. \ref{Eq:totflux}. The left-hand panel of Fig. \ref{Figure:balmeraccretion} shows an example of a photospheric synthetic spectra modified by the contribution of accretion. This contribution causes the total flux to show a filled-in Balmer discontinuity, compared to that of the undisturbed photosphere.

For a given set of stellar and accreting parameters, the modelled excess in the Balmer discontinuity is defined as $\Delta$D$_B$ =  $(U-B)$$_{phot}$ -- $(U-B)$$_{total}$. This is computed by subtracting the $U$-$B$ colour of the synthetic total flux, which accounts for the influence of accretion, from the corresponding Kurucz photospheric colour, which reflects the naked photosphere. The fluxes were convolved with the $U$ and $B$ transmission curves provided by the Nordic Optical Telescope (left panel of Fig. \ref{Figure:balmeraccretion}), since the photometric observations were carried out with the instrument TurPol on that telescope \citep{Oudmaijer01}. That the photometric points are slightly displaced from the spectra is caused by the strong gradients in the fluxes along the transmission bands, specially in $U$. The difference between the Kurucz colours computed in this way and the intrinsic colours in \citet{KenyonHartmann95} is $\leq$ 0.05 magnitudes for the stellar parameters considered here, being typically around 0.01 magnitudes. This accuracy is enough for our purposes and means that the modelled excesses can later be compared to the observed ones, described below. The right-hand panel of Fig. \ref{Figure:balmeraccretion} shows a representative subset of the $\Delta$D$_B$--$\dot{M}_{\rm acc}$ calibrations we have modelled. The curves rise with the corresponding filling factors, following Eq. \ref{Eq:condition2}, and are limited by $f$ $<$ 1. There is a strong dependence on the stellar temperature and surface gravity that must be considered when a mass accretion rate is associated to a given $\Delta$D$_B$ value.

\begin{figure}
\centering
 \includegraphics[width=90mm,clip=true]{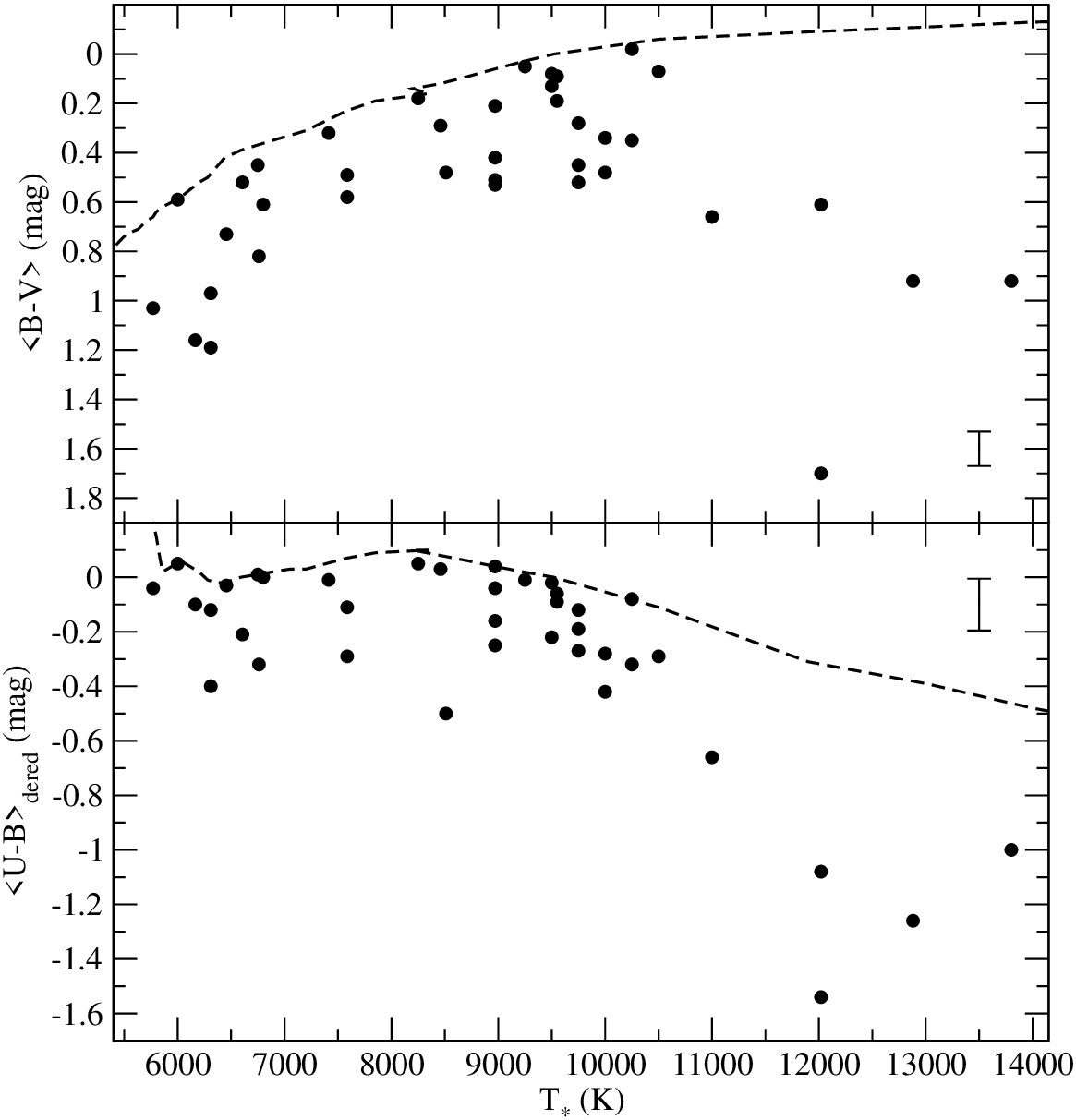}
\caption{Observed mean colours in $B-V$ (top) and $U-B$ (bottom) as a function of the stellar temperature. The colours in the bottom panel are deredenned using the excesses derived from the top panel. The dashed lines indicate the corresponding intrinsic colours from \citet{KenyonHartmann95}. The typical error bars considering the photometric uncertainties are shown.}
\label{Figure:T_Balmerjump}
\end{figure}

From the observational perspective, the mean excess in the Balmer discontinuity for a given star is $<$$\Delta$D$_B$$>$ =  $(U-B)$$_0$ -- $<$$U-B$$>$$_{dered}$, where $(U-B)$$_0$ is the intrinsic colour from \citet{KenyonHartmann95}, and $<$$U-B$$>$$_{dered}$ the dereddened mean colour, obtained from the observations in \citet{Oudmaijer01}. The dereddening was applied using the $<$E(B--V)$>$ values in Table \ref{Table:sample}, R$_{\rm V}$ = 5, and the same extinction law mentioned in Sect. \ref{Section:Sample}. The top panel in Fig. \ref{Figure:T_Balmerjump} shows the mean observed and intrinsic $B-V$ colours from which the dereddening was applied. The hottest stars in our sample are the most reddened. The bottom panel shows the $<$$U-B$$>$$_{dered}$ and $(U-B)$$_0$ colours from which the $<$$\Delta$D$_B$$>$ values are obtained. The hottest stars are also the ones showing the largest Balmer excesses. It worth mentioning that a slight variation in the adopted value for the stellar temperature translates into a strong change in $<$$\Delta$D$_B$$>$, for T$_*$ $\lesssim$ 6400 K. This is caused by the steep dependence of both the $U-B$ and $B-V$ intrinsic colours on that range of stellar temperatures \citep[e.g. differences up to 0.2 magnitudes for variations of $\pm$ 150 K, see][]{KenyonHartmann95}. In turn, this could significantly change the accretion rate estimate (e.g. a variation of 0.2 magnitudes in the Balmer excess translates into a change up to two orders of magnitude in $\dot{M}_{\rm acc}$ for T$_*$ = 6500 K; see right panel of Fig. \ref{Figure:balmeraccretion}). The dependence of the intrinsic colours on T$_*$ is smoother for almost all the stars studied here (Fig. \ref{Figure:T_Balmerjump}). The typical uncertainty for $<$$\Delta$D$_B$$>$ is between 0.05 and 0.1 magnitudes in our sample. This was estimated by considering both uncertainties in the photospheric colours, from variations in T$_{*}$ of $\pm$ 150 K, and the photometric uncertainties, which are the dominant contribution for most stars. A typical uncertainty of 0.07 magnitudes is adopted in this paper for $<$$\Delta$D$_B$$>$, but that strictly depends on each star and could increase for the coldest objects in the sample.

Individual calibrations, similar to the examples in the right-hand panel of Fig. \ref{Figure:balmeraccretion}, were constructed for each star from the corresponding M$_{*}$, T$_{*}$, R$_{*}$, and log g parameters in Table \ref{Table:sample}. In this way, a mass accretion rate was assigned to the modelled Balmer excess that matches the observed one. Then $L$$_{\rm acc}$ was obtained from $\dot{M}_{\rm acc}$, M$_{*}$, and R$_{*}$. This method does not need distance calibrations since it is based on colour excesses. In addition, the stellar parameters of most stars in the sample were obtained from a distance-independent method \citep{Montesinos08}. However, distance uncertainties are indirectly introduced through R$_{*}$ -and log g- when this was derived from a value of L$_{*}$ based on a distance \citep[e.g.][]{Manoj06}.   

\subsection{Results}
\label{section:discuss}

Table \ref{Table:accretion} shows the observed (mean) excesses in the Balmer discontinuity and the accretion rates and best model parameters that reproduce them. The uncertainties for $\dot{M}_{\rm acc}$ were estimated by varying $\Delta$D$_B$ by $\pm$0.07 magnitudes in the models, which is the typical uncertainty for the observed excess (Sect. \ref{section:descriptmodel}). Upper limits are provided when $<$$\Delta$D$_B$$>$ is lower than that value. The uncertainties for $L$$_{\rm acc}$ also consider the typical error in the M$_{*}$/R$_{*}$ ratios. In addition to the sources of error discussed in these sections, we refer the reader to the broad analysis in \citet{Herczeg08} on how different uncertainties of the stellar and model assumptions could affect the accretion rates estimates.

\begin{table}
\centering
\renewcommand\tabcolsep{3.2pt}
\caption{Observed mean excess in the Balmer discontinuity, accretion rates, and model parameters.}
\label{Table:accretion}
\begin{tabular}{lrrrrrr}
\hline\hline
Star&$<$$\Delta$D$_B$$>$&log $\dot{M}_{\rm acc}$&log $L$$_{\rm acc}$&$R$$_i$&$T$$_{col}$&$f$\\
 &(mag)& [M$_{\sun}$ yr$^{-1}$]&[L$_{\sun}$]&(R$_{*}$)&(K)&($\%$) \\
\hline
HD 31648   & 0.05  &$<$-7.23	     &$<$0.20          &2.5&12215&1.1 \\
HD 34282   & 0.06  &$<$-8.30	     &$<$-0.40         &2.5&12695&2.2 \\
HD 34700   & 0.00  &$<$-8.30	     &$<$-1.05         &2.5&11730&0.02\\
HD 58647   & 0.18  &   -4.84$\pm$0.22&   2.47 $\pm$0.23&2.1&13140&12  \\
HD 141569  & 0.09  &   -6.89$\pm$0.40&   0.70 $\pm$0.40&1.5&12695&3.6 \\
HD 142666  & 0.18  &   -6.73$\pm$0.26&   0.69 $\pm$0.27&2.5&12030&3.2 \\
HD 144432  & 0.06  &$<$-7.22	     &$<$0.21          &2.5&11990&1.1 \\
HD 150193  & 0.29  &   -6.12$\pm$0.14&   1.33 $\pm$0.15&2.5&12460&13  \\
HD 163296  & 0.02  &$<$-7.52	     &$<$-0.03         &2.2&12570&0.61\\
HD 179218  & 0.02  &$<$-7.30	     &$<$0.14          &2.5&12670&0.60\\
HD 190073  & 0.22  &   -5.00$\pm$0.25&   2.29 $\pm$0.26&2.5&12670&12  \\
AS 442     & 0.48  &   -5.08$\pm$0.11&   2.37 $\pm$0.12&2.5&13405&56  \\
VX Cas     & 0.22  &   -6.44$\pm$0.22&   1.16 $\pm$0.23&2.0&12895&13  \\
BH Cep     & 0.01  &$<$-8.30	     &$<$-0.94         &2.4&11800&0.07\\
BO Cep     & 0.21  &   -6.93$\pm$0.28&   0.45 $\pm$0.29&2.5&11825&2.8 \\
SV Cep     & 0.24  &   -6.30$\pm$0.20&   1.30 $\pm$0.21&1.8&13015&14  \\
V1686 Cyg  & 0.12  &   -5.23$\pm$0.41&   1.66 $\pm$0.41&2.5&11755&0.87\\
R Mon      & 0.76  &(?)    	     &(?)              &...&...	 &... \\
VY Mon     & 1.22  &(?)    	     &(?)              &...&...	 &... \\
51 Oph	   & 0.00  &$<$-7.00	     &$<$0.37          &1.3&13015&0.11\\
KK Oph	   & 0.36  &   -5.85$\pm$0.15&   1.51 $\pm$0.16&1.6&12030&9.4 \\
T Ori      & 0.09  &   -6.58$\pm$0.40&   0.90 $\pm$0.40&1.8&12780&3.7 \\
BF Ori     & 0.00  &$<$-8.00	     &$<$-0.60         &2.5&12460&0.09\\
CO Ori     & 0.38  &   -5.20$\pm$0.18&   1.93 $\pm$0.19&2.5&11775&4.5 \\
HK Ori	   & 0.57  &   -5.24$\pm$0.12&   2.13 $\pm$0.13&2.5&12300&31  \\
NV Ori     & 0.00  &$<$-8.30	     &$<$-0.99         &2.5&11850&0.03\\
RY Ori     & 0.11  &   -6.65$\pm$0.33&   0.59 $\pm$0.33&2.5&11775&0.75\\
UX Ori	   & 0.05  &$<$-6.77	     &$<$0.63          &1.5&12285&1.1 \\
V346 Ori   & 0.25  &   -5.99$\pm$0.17&   1.47 $\pm$0.18&2.5&12780&15  \\
V350 Ori   & 0.20  &   -6.66$\pm$0.24&   0.82 $\pm$0.25&2.5&12460&5.0 \\
XY Per     & 0.16  &   -5.86$\pm$0.20&   1.57 $\pm$0.21&1.5&12780&7.5 \\
VV Ser	   & 0.54  &(?)  	     &(?)              &1.7&...  &...  \\
CQ Tau	   & 0.02  &$<$-8.30	     &$<$-0.84         &2.5&11860&0.21\\
RR Tau	   & 0.37  &   -4.11$\pm$0.16&   3.18 $\pm$0.17&1.3&12895&27  \\
RY Tau     & 0.24  &   -7.65$\pm$0.17&   -0.22$\pm$0.18&2.5&11700&0.98\\
PX Vul     & 0.33  &   -6.72$\pm$0.16&   0.73 $\pm$0.17&2.5&11850&7.2 \\
WW Vul	   & 0.08  &   -6.38$\pm$0.70&   1.05 $\pm$0.70&1.5&12460&2.7 \\
LkHa 234   & 0.88  &(?)    	     &(?)     	       &...&...	 &... \\
\hline

\end{tabular}
\begin{minipage}{9cm}

  \textbf{Notes.} A typical uncertainty of 0.07 magnitudes is adopted for all $<$$\Delta$D$_B$$>$ values. The symbol (?) refers to the four objects for which strong difficulties are found to reproduce their Balmer excesses from our model. 
\end{minipage}
\end{table} 

The high $<$$\Delta$D$_B$$>$ value shown by \object{VV Ser} cannot be reproduced using the typical value $\curf$ = 10$^{12}$ erg cm$^{-2}$ s$^{-1}$. Increasing $\curf$ by one order of magnitude provides $\dot{M}_{\rm acc}$  $\sim$ 10$^{-4}$ M$_{\sun}$ yr$^{-1}$, but the filling factor is almost 100$\%$. Further increases in $\curf$ produce lower filling factors, but the corresponding accretion rates also increase. Another possibility is to increase the disk truncation radius to a value significantly higher than the co-rotation one ($\sim$ 1.7 R$_{*}$ for \object{VV Ser}). For $R$$_{i}$ = 5.5 R$_{*}$ and $\curf$ = 10$^{13}$ erg cm$^{-2}$ s$^{-1}$, the Balmer excess of \object{VV Ser} can be reproduced from $\dot{M}_{\rm acc}$  $\sim$ 5$\times$10$^{-5}$ M$_{\sun}$ yr$^{-1}$ (f = 0.99). Similarly, we find it difficult to reproduce the strong Balmer excesses shown by \object{R Mon}, \object{VY Mon}, and \object{LkHa 234} from the model used in this work. For the highest accretion rates possible ($\sim$ 10$^{-2}$--10$^{-1}$ M$_{\sun}$ yr$^{-1}$, with $\curf$ $>$$>$ 10$^{12}$ erg cm$^{-2}$ s$^{-1}$ and $f$ $\sim$ 1), their modelled spectra are featureless and provide Balmer excesses that are $\sim$ 50$\%$ lower than observed. Again, a combination of large disk truncation radius, well above the typical 2.5 R$_{*}$ for HAeBe stars, $\curf$ $\gg$ 10$^{12}$ erg cm$^{-2}$ s$^{-1}$, and f $\sim$ 1 could reproduce the Balmer excess of these objects from lower accretion rates ($\sim$ 10$^{-5}$ M$_{\sun}$ yr$^{-1}$). However, the use of those values would be arbitrary and not supported by theory or observations (see MDCH04). In particular, a disk truncation radius larger than the co-rotation one is in principle not possible in the MA scenario \citep[see e.g.][]{Shu94}. Temperatures and extinction properties that are different than the ones used in this work would make their Balmer excesses in Table \ref{Table:accretion} wrong, and therefore useless in our fit attempts. However, It is noted that the four problematic objects not only show the highest $<$$\Delta$D$_B$$>$ values in our sample, but are also the hottest (T$_{*}$ $>$ 12000K, i.e. spectral-types earlier than $\sim$ B8) and most heavily extincted sources. There is consensus in the literature that these objects are early-type HAeBe stars \citep[see e.g. the compilation of spectral types in Table 6 of][]{Mora01}. Our previous results, indicating that the variability of the $H\alpha$ line is clearly lower for the massive early-type HAeBe stars, already pointed to different physical processes operating in their circumstellar environments (see Paper I). The lack of success in reproducing the Balmer excesses could be pointing to a change in the accretion paradigm for the HBe regime (see also Sect. \ref{section:W10}).  No result is included in Table \ref{Table:accretion} for the four objects discussed, which would require specific analysis. The calibrations with the line luminosities included in Sect. \ref{section:actracers} and the stellar parameters used in this work provide mass accretion rates (in M$_{\sun}$ yr$^{-1}$) of $\sim$ 10$^{-5}$--6$\times$10$^{-6}$ for \object{VV Ser}, 3$\times$10$^{-4}$--3$\times$10$^{-5}$ for \object{R Mon}, 3$\times$10$^{-3}$--7$\times$10$^{-6}$ for \object{VY Mon}, and 6$\times$10$^{-5}$--8$\times$10$^{-5}$ for \object{LkHa 234}. These estimates must be viewed with caution, given the discussion above and their being based on relations derived for colder stars. 

The typical (median) accretion rate for the remaining objects is 2 $\times$ 10$^{-7}$ M$_{\sun}$ yr$^{-1}$, 36$\%$ of the stars showing $\dot{M}_{\rm acc}$ $\leq$ 10$^{-7}$ M$_{\sun}$ yr$^{-1}$, 35$\%$ between 10$^{-7}$ and 10$^{-6}$ M$_{\sun}$ yr$^{-1}$, and 29$\%$ larger than 10$^{-6}$ M$_{\sun}$ yr$^{-1}$. This distribution is practically the same as predicted from Fig. 8 of MDCH04. 

\begin{figure*}
\centering
 \includegraphics[width=180mm,clip=true]{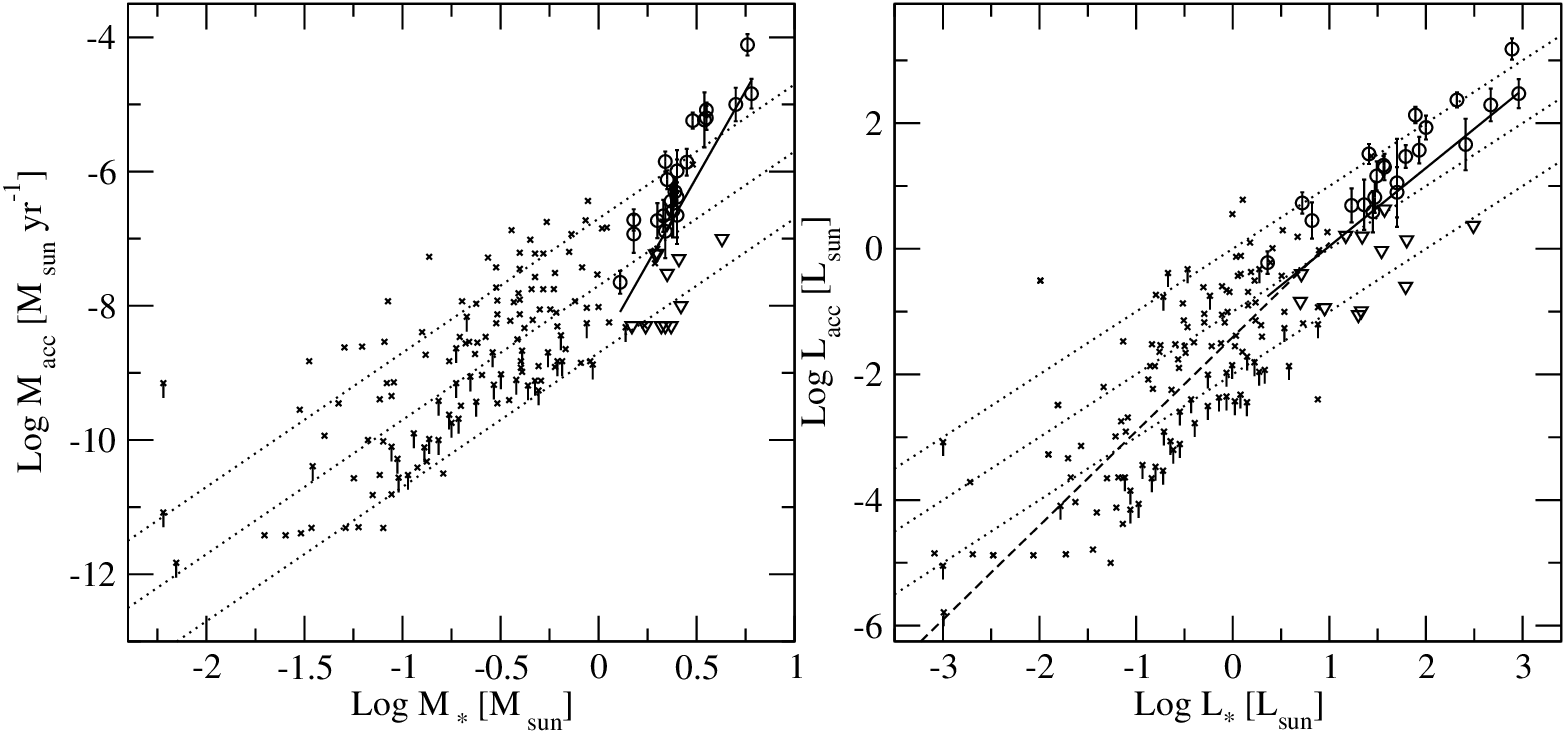}
\caption{Mass accretion rate vs stellar mass (left) and accretion luminosity vs stellar luminosity (right). Our data are indicated with circles, and triangles for the upper limits on the accretion rates. Crosses are data from the literature (those with vertical bars are upper limits), including low-mass stars from different star-forming regions \citep[see][and references therein]{Natta06}. Dotted lines indicate $\dot{M}_{\rm acc}$ $\propto$ M$_{*}^{2}$ $\pm$ 1 dex (left) and $L$$_{\rm acc}$/L$_{*}$ = 0.01, 0.1, 1 (right). The dashed line on the right side is the best fit for low-mass stars ($L$$_{\rm acc}$ $\propto$ L$_{*}^{1.5}$), and the solid lines are the best fits for our sample ($\dot{M}_{\rm acc}$ $\propto$ M$_{*}^{5}$ and $L$$_{\rm acc}$ $\propto$ L$_{*}^{1.2}$).}
\label{Figure:miaccretion}
\end{figure*}

\begin{figure}
\centering
 \includegraphics[width=90mm,clip=true]{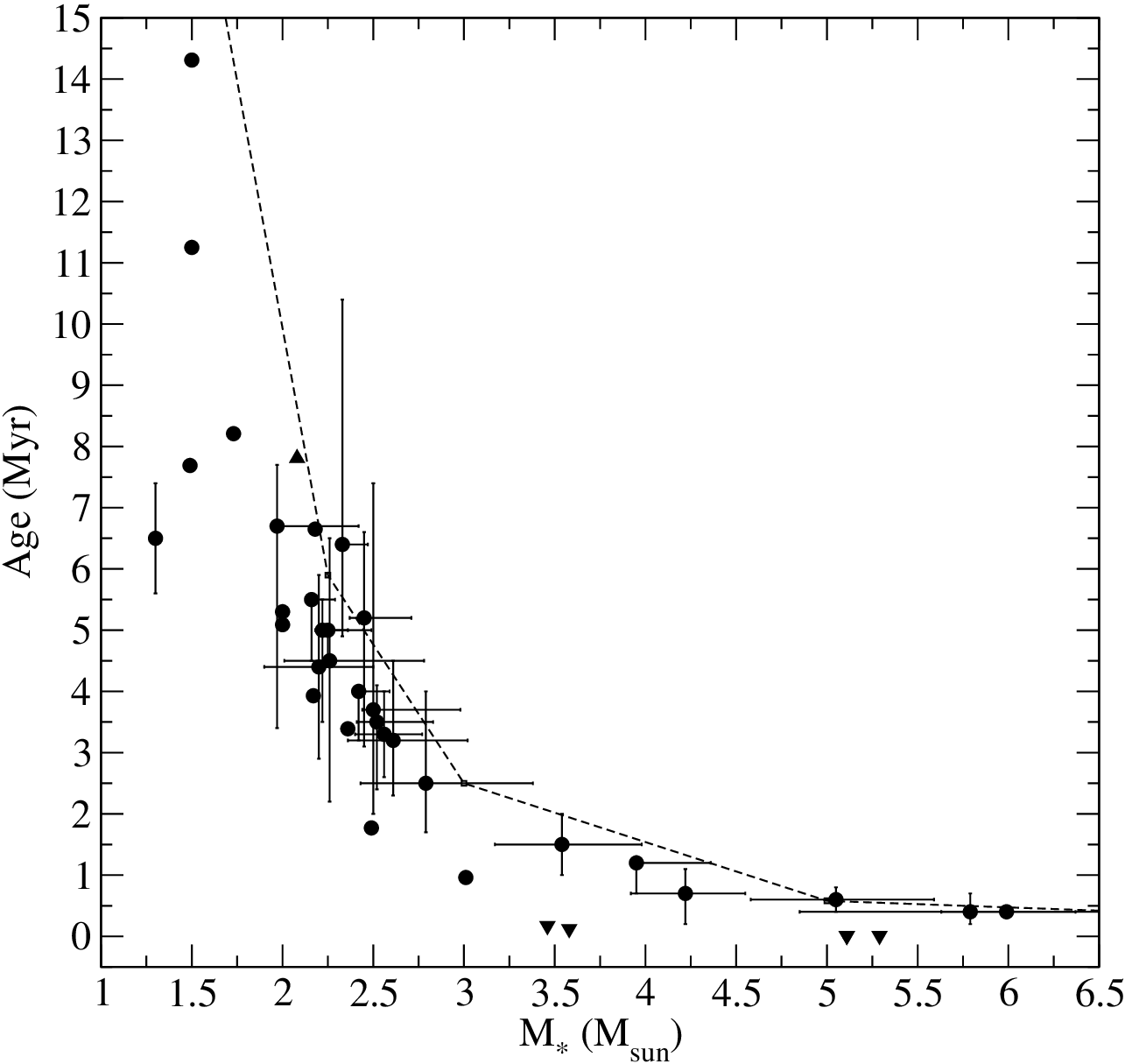}
\caption{Stellar age versus stellar mass for the stars in the sample. Error bars are plotted when provided in the references of Table \ref{Table:sample}. The dashed line indicates the theoretical time to reach the main sequence \citep{Tayler94}.}
\label{Figure:M_age}
\end{figure}

Figure \ref{Figure:miaccretion} (left panel) shows the mass accretion rates against the stellar masses. The stars in our sample follow a steeper trend than the lower mass objects, for which $\dot{M}_{\rm acc}$ $\propto$ M$_{*}^{2}$; \citep[see e.g.][]{Muzerolle03,Muzerolle05,Mohanty05,Natta06}. We find $\dot{M}_{\rm acc}$ $\propto$ M$_{*}^{5.2}$ for our sample (Pearson's correlation coefficient 0.74). The exponent decreases to 4.6 if the upper limits for the accretion rates are discarded (Pearson's correlation coefficient 0.91). Our data is therefore reasonably well fitted to $\dot{M}_{\rm acc}$ $\propto$ M$_{*}^{5}$. This trend is driven by the most massive sources and is most probably caused by the mass-age distribution characterizing HAeBe stars \citep[see e.g.][]{vanboekel05,GarciaLopez06}. This distribution is shown in Fig. \ref{Figure:M_age} for our sample. The stars with M$_{*}$ $\geq$ 2.5M$_{\sun}$ are younger than $\sim$ 2 Myr, and the objects with lower stellar masses are older. Following \citet{vanboekel05}, the lack of “old” HAeBes around 2.5-6 M$_{\odot}$ stars is likely caused by their faster evolution to the MS, which is supported by observations \citep{Roccatagliatta11}. The scarcity of “young” HAeBes around 1-2.5M$_{\odot}$ stars is most probably explained by the fact that they become optically visible later in their evolution. The massive young stars in our sample show the strongest accretion rates, as expected. The mass accretion rates of the less massive older HAeBe stars are orders of magnitude lower. These facts would make the $\dot{M}_{\rm acc}$--M$_{*}$ trend very steep. 

The right-hand panel in Fig. \ref{Figure:miaccretion} shows the accretion luminosities against the stellar ones. The stars in our sample have 0.01L$_{*}$ $\leq$ $L$$_{\rm acc}$ $\leq$ L$_{*}$. Both limits are similar to those reported for lower mass stars \citep{Clarke06,Tilling08}. The lower limit is the observational detection threshold for $L$$_{\rm acc}$. However, although the trend between $L$$_{\rm acc}$ and L$_{*}$ shown by low-mass stars \citep[$L$$_{\rm acc}$ $\propto$ L$_{*}^{1.5}$; see e.g.][]{Tilling08} is roughly followed by our sample, our best fit ($L$$_{\rm acc}$ $\propto$ L$_{*}^{1.2}$) could indicate a decrease in the exponent for the intermediate-mass regime. In fact, a direct visual inspection of the right-hand panel in Fig. \ref{Figure:miaccretion} reveals that there is an apparent decline in the slope of the lower envelope of the $L$$_{\rm acc}$--L$_{*}$ trend. That decrease could start at lower stellar luminosities than those analysed in this work, around  L$_{*}$ = L$_{\sun}$.

\subsubsection{Comparison with previous results}
\label{section:prevresults}
\begin{figure}
\centering
 \includegraphics[width=90mm,clip=true]{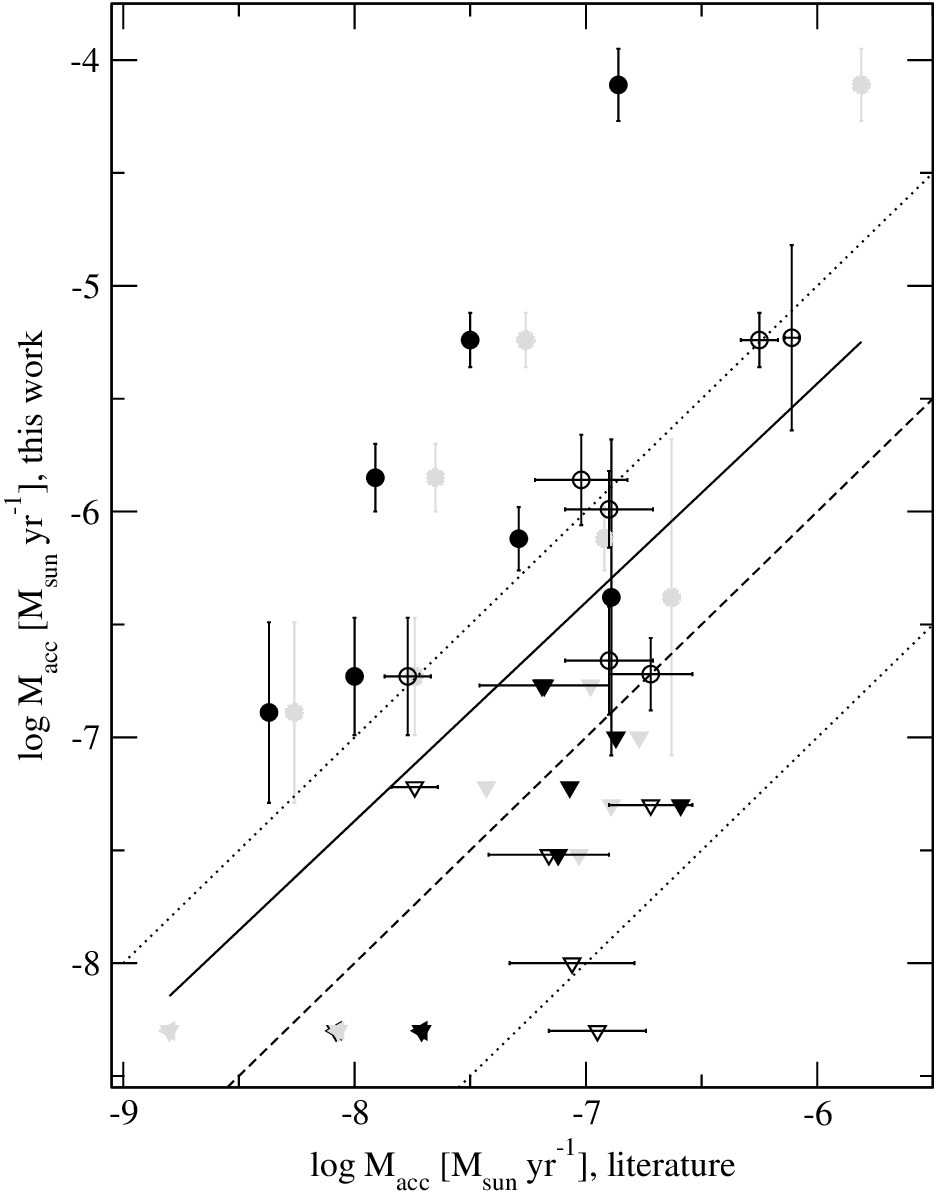}
\caption{Mass accretion rates derived in this work against those derived by \citet{GarciaLopez06} (filled symbols) and \citet{DonBrit11} (open symbols). The dashed line indicates equal values and the dotted lines $\pm$1 dex. The typical minimum uncertainty for the estimates in \citet{GarciaLopez06} is $\pm$ 0.5 dex. Triangles are upper limits. Grey symbols represent the estimates using the method in \citet{GarciaLopez06} from our distances and stellar parameters, and the solid line is the best fit.}
\label{Figure:accr_comp}
\end{figure}

\citet{GarciaLopez06} obtained accretion luminosities for a sample of HAeBe stars from the empirical calibration with the Br$\gamma$ luminosity for intermediate-mass T Tauri stars \citep[log ($L$$_{\rm acc}$/L$_{\sun}$) = 0.9$\times$log (L$_{{\rm Br}\gamma}$/L$_{\sun}$) + 2.9;][]{Calvet04}. The non-photospheric Br$\gamma$ equivalent widths were transformed to luminosities from the $K$ magnitudes and distances in their Table 1. The mass accretion rates were then derived from their values for M$_{*}$ and R$_{*}$, using the same $L$$_{\rm acc}$--$\dot{M}_{\rm acc}$ expression as in this work (Sect. \ref{section:descriptmodel}). Figure \ref{Figure:accr_comp} shows our mass accretion rates against those from \citet{GarciaLopez06} for the objects in common. The difference between both determinations decreases once the Br$\gamma$ luminosities are rescaled to our distances (see Table \ref{Table:sample}) and  the correspoding accretion luminosities are transformed into mass accretion rates from our values for M$_{*}$ and R$_{*}$. The grey symbols in Fig. \ref{Figure:accr_comp} are obtained in this way, representing the estimates from the method in \citet{GarciaLopez06} rescaled to our values for d, M$_{*}$, and R$_{*}$. The new estimates show a linear correlation with our results (Pearson's coefficient 0.65) with slope 1.0 $\pm$ 0.3, and displaced in the ordinate axis by a factor $\sim$ 0.4. This agrees with the analysis in the following section. It must be noted that the stellar temperature for several objects in \citet{GarciaLopez06} falls well above the upper limit from which the Br$\gamma$ relation in \citet{Calvet04} was obtained -around 6100 K-.

\citet{DonBrit11} have derived mass accretion rates for a sample of HAeBe stars from the calibration with the Balmer excess provided by \citet{Muzerolle04}. The Balmer excesses were obtained fitting both Kurucz and standard main-sequence spectra to UV-optical spectra. Figure \ref{Figure:accr_comp} shows our mass accretion rates versus the ones from \citet{DonBrit11} for the stars included in both samples. Both estimates are equal within $\pm$ 1 dex for almost all objects in common. Variability could account for differences up to $\sim$ 0.5 dex (see Sect. \ref{section:accretionvar}), but the scatter can be explained also from other factors. First, the method used to derive the Balmer excess is different in both papers. Second, the stellar parameters differ significantly for several stars (e.g. $\Delta$T$_{*}$ $>$ 10000 K for \object{V1686 Cyg}). Finally, the calibration used by \citet{DonBrit11} was in many cases applied to stars with stellar parameters significantly different than those from which the calibration was made \citep[spectral type A2, M$_{*}$ = 3 M$_{\odot}$ and R$_{*}$ = 3 R$_{\odot}$][]{Muzerolle04}. As discussed in Sect. \ref{section:descriptmodel}, the $\Delta$D$_B$--$\dot{M}_{\rm acc}$ calibration differs strongly depending on the stellar parameters considered (right panel in Fig. \ref{Figure:balmeraccretion}).

The mass accretion rate provided for \object{CO Ori} in \citet{Calvet04} was derived from a similar method to the one used in this paper, but it was a factor $\sim$ 70 lower. This is explained mainly from differences in the stellar parameters adopted. Both L$_{*}$ and M$_*$ are higher in the present work by a factor $\sim$ 45 and 1.4, respectively. From L$_{*}$ and T$_{*}$ (see below), our R$_*$ is a factor 2 larger. The values adopted for R$_{*}$ and M$_*$ affect those for the filling factor and the blackbody temperature representing the accretion shock. The values for the surface gravities and effective temperatures are log g = 3.6 against our 3.2, and T$_*$ = 6030 K against our 6310 K. This difference of almost 300 K in that range of stellar temperatures, along with the slight differences in the $U-B$ and $B-V$ colours -around 0.05 magnitudes-, translates into a significant change in the Balmer excess -0.24 magnitudes from the data in \citet{Calvet04} against our 0.38 magnitudes; see the discussion in Sect. \ref{section:descriptmodel}-. The higher mass accretion rate we provide agrees with the one expected for a star that here is considered hotter, with a lower value for log g and with more Balmer excess (see right panel in Fig. \ref{Figure:balmeraccretion}), but our model yields the same $\dot{M}_{\rm acc}$ $\sim$ 10$^{-7}$--10$^{-8}$ M$_{\sun}$ yr$^{-1}$ for \object{CO Ori} using the parameters in \citet{Calvet04}. The different extinction treatment is also relevant, but other factors such as the use of different values for $\curf$ and $R$$_i$ have less influence. The mass accretion rate for the other object in common with \citet{Calvet04}, \object{RY Tau}, is a factor $\sim$ 3 higher than our estimate, but the values overlap ($\sim$ 10$^{-8}$ M$_{\sun}$ yr$^{-1}$) if considering the uncertainties from both papers. Apart from the practically equal values for $U-B$ and $B-V$, again within 0.05 magnitudes, the stellar parameters adopted in both works are more similar than for \object{CO Ori}.

Given the relevance of the stellar characterization to derive the accretion rates, we notice that most of the stellar parameters in Table \ref{Table:sample} were carefully obtained by \citet{Montesinos08} in a paper specifically devoted to that task, which takes much of the photometry used in this work into account.
 
\section{Accretion tracers}
\label{section:actracers}
\begin{figure*}
\centering
 \includegraphics[width=180mm,clip=true]{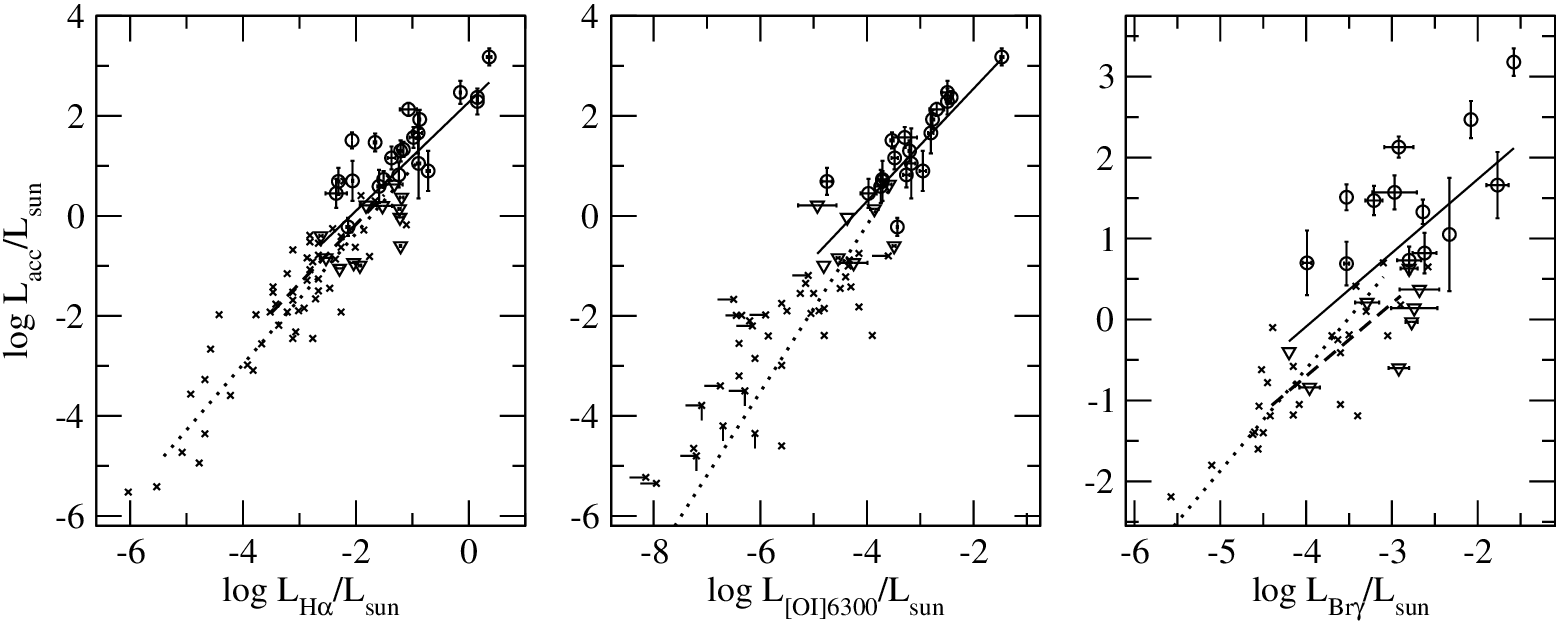}
\caption{Accretion luminosity against the H$\alpha$ (left), [\ion{O}{i}]6300 (middle) and Br$\gamma$ (right) luminosities . Our results are represented with circles and triangles for the upper limits on the accretion rates. Our accretion and Br$\gamma$ luminosities were not derived from simultaneous measurements. Error bars for the line luminosities indicate the uncertainties for the non-variable objects and variability amplitudes for the variable ones. Our best fits are plotted with solid lines. Data for lower mass stars (crosses) were taken from \citet{Fang09,Herczeg08}, and \citet{Calvet04}, for the left, middle, and right panels, respectively. Upper limits are indicated with bars, when provided. Dotted and dashed lines are the empirical calibrations for H$\alpha$ and Br$\gamma$ in low and intermediate-mass CTT, respectively (see text).}
\label{Figure:emcor}
\end{figure*}

Figure \ref{Figure:emcor} shows our accretion luminosities in Table \ref{Table:accretion} against the H$\alpha$, [\ion{O}{i}]6300 and Br$\gamma$ luminosities in Table \ref{Table:sample}. The best least-squares fits provide the following empirical calibrations for our sample (1 $\leq$ M$_{*}$/M$_{\sun}$ $\leq$ 6; 6000 $\leq$ T$_{*}$(K) $\leq$ 11000):
\begin{equation}
\label{Eq:coralpha}
\log \left(\frac{{\rm L}_{\rm acc}}{{\rm L}_{\sun}}\right) = 
2.28(\pm0.25) + 1.09(\pm0.16)\times\log \left(\frac{{\rm L}_{{\rm H}\alpha}}{{\rm L}_{\sun}}\right)
\end{equation}
\begin{equation}
\label{Eq:corOI}
\log \left(\frac{{\rm L}_{\rm acc}}{{\rm L}_{\sun}}\right) = 
4.80(\pm0.50) + 1.13(\pm0.14)\times\log \left(\frac{{\rm L}_{{\rm [OI]}6300}}{{\rm L}_{\sun}}\right)
\end{equation}
\begin{equation}
\label{Eq:corgamma}
\log \left(\frac{{\rm L}_{\rm acc}}{{\rm L}_{\sun}}\right) = 
3.55(\pm0.80) + 0.91(\pm0.27)\times\log \left(\frac{{\rm L}_{{\rm Br}\gamma}}{{\rm L}_{\sun}}\right).
\end{equation}

These empirical expressions are very similar to those provided for low-mass CTT stars \citep{Muzerolle98a,Dahm08,Herczeg08}:
\begin{equation}
\label{Eq:coralphalow}
\log \left(\frac{{\rm L}_{\rm acc}}{{\rm L}_{\sun}}\right) = 
2.27(\pm0.70) + 1.31(\pm0.16)\times\log \left(\frac{{\rm L}_{{\rm H}\alpha}}{{\rm L}_{\sun}}\right)
\end{equation}
\begin{equation}
\label{Eq:corOIlow}
\log \left(\frac{{\rm L}_{\rm acc}}{{\rm L}_{\sun}}\right) = 
6.50(\pm2.18) + 1.67(\pm0.28)\times\log \left(\frac{{\rm L}_{{\rm [OI]}6300}}{{\rm L}_{\sun}}\right)
\end{equation}
\begin{equation}
\label{Eq:corgammalow}
\log \left(\frac{{\rm L}_{\rm acc}}{{\rm L}_{\sun}}\right) = 
4.43(\pm0.79) + 1.26(\pm0.19)\times\log \left(\frac{{\rm L}_{{\rm Br}\gamma}}{{\rm L}_{\sun}}\right).
\end{equation}

There is an apparent decrease in the slope of the calibrations for the HAeBe regime, but firm conclusions can only be stated for the [\ion{O}{i}]6300 line; the slope in Eq. \ref{Eq:corOI} is shallower than that in Eq. \ref{Eq:corOIlow} above the uncertainties. However, the H$\alpha$ and Br$\gamma$ slopes for intermediate-mass T-Tauri stars are 1.18$\pm$0.26 \citep{Dahm08} and 0.9 \citep{Calvet04}, respectively, i.e. closer to our estimates \citep[see also][]{DonBrit11}. This could be indicating that the slopes of the empirical calibrations decrease for higher stellar masses. The scatter for a given line luminosity is comparable for classical T-Tauri and HAeBe stars. The slightly higher dispersion for the Br$\gamma$ line could likely be diminished using simultaneous measurements for the accretion and line luminosities.

As introduced in Sect. \ref{Section:Sample}, a different value for the total-to-selective extinction ratio would influence only the Balmer excesses of the most heavily extincted sources in our sample. These tend to be the early-type HAeBe stars, which are the strongest accretors (see Fig. \ref{Figure:T_Balmerjump} and Sect. \ref{section:discuss}). Using R$_{\rm V}$ $<$ 5 would make their accretion rate estimates lower. Therefore, the $L$$_{\rm acc}$--L$_{*}$ and $L$$_{\rm acc}$--L$_{line}$ trends for our sample in the right-hand panel of Fig. \ref{Figure:miaccretion} and in Fig. \ref{Figure:emcor} would flatten slightly. This would enhance the difference between the relations for the low-mass and the HAeBe regimes. In addition, The $\dot{M}_{\rm acc}$--M$_{*}$ relation in the left-hand panel of Fig. \ref{Figure:miaccretion} would show a flatter slope, but the trend would be still steeper than for low-mass stars, even when using R$_{\rm V}$ = 3.1.

\subsection{Accretion and the H$\alpha$ 10$\%$ width}
\label{section:W10}
\begin{figure}
\centering
 \includegraphics[width=90mm,clip=true]{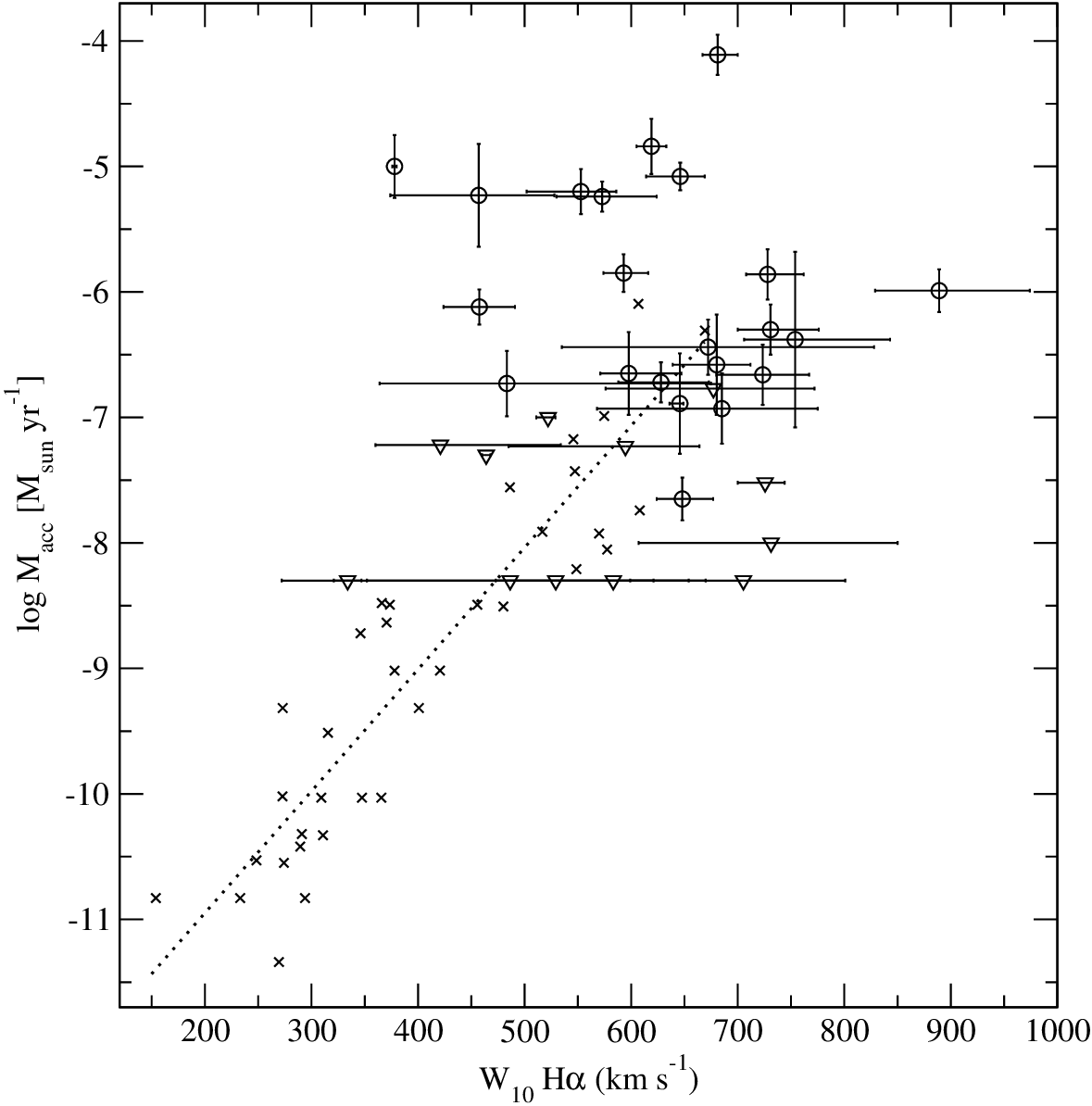}
\caption{Mass accretion rate against the H$\alpha$ 10$\%$ width. Our results are represented by circles, and triangles for the upper limits on the accretion rates. Error bars for the line widths indicate the uncertainties for the non-variable objects and variability amplitudes for the variable ones. Data for lower mass stars are also included \citep[crosses; from the compilation of][]{Natta04}, as well as the empirical expression relating both parameters in that stellar regime; log $\dot{M}_{\rm acc}$ $\sim$ -12.9 + 9.7$\times$ 10$^{-3}$W$_{10}$(H$\alpha$) \citep[dotted line;][]{Natta04}.}
\label{Figure:W10_Macc}
\end{figure}

The H$\alpha$ width at 10$\%$ of peak intensity is widely used as an empirical accretion tracer for low-mass stars \citep{White03,Natta04,Jayawardhana06}. Figure \ref{Figure:W10_Macc} shows the mass accretion rates against the H$\alpha$ 10$\%$ widths. The trend followed by low-mass objects \citep{Natta04} breaks for HAeBe stars, whose $<$W$_{10}$(H$\alpha$)$>$ values show no correlation with log $\dot{M}_{\rm acc}$. This was expected since the mean values and the relative variabilities of the H$\alpha$ equivalent and 10$\%$ widths are not correlated in our sample (see Paper I). We conclude that $<$W$_{10}$(H$\alpha$)$>$ can usually not be used as a direct tracer of accretion for HAeBe stars, unlike in the lower mass regime. We argue in the following that one main reason for this discrepancy is that the typically high rotation rates of massive stars \citep{Finkenzeller85,Mora01} influence their $<$W$_{10}$(H$\alpha$)$>$ values. We suggest that this influence can be interpreted in the context of the MA scenario.

\begin{figure}
\centering
 \includegraphics[width=90mm,clip=true]{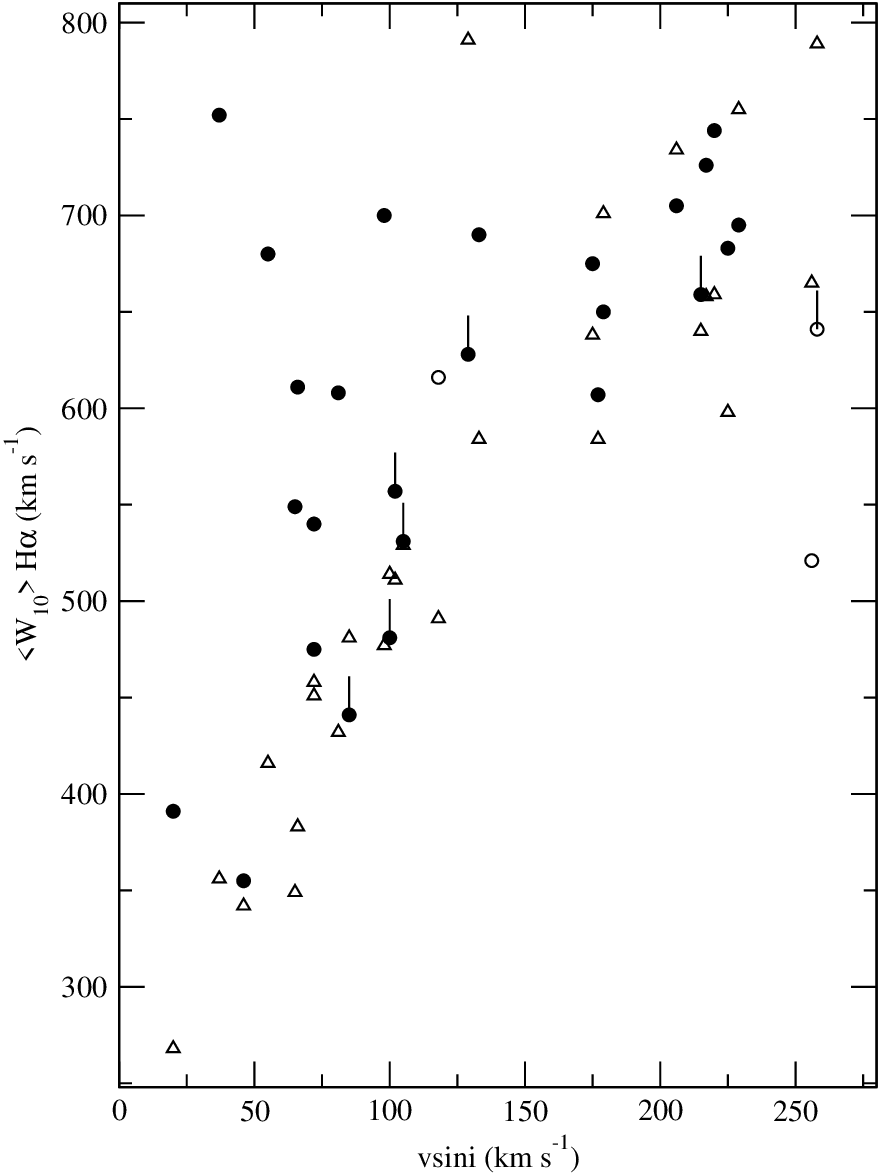}
\caption{H$\alpha$ mean line width at 10$\%$ of peak
  intensity (from Paper I) vs the projected rotational velocities (circles). Lower limits for $<$W$_{10}$(H$\alpha$)$>$ are marked with vertical bars. Open circles indicate HBe stars -\object{51 Oph} the object with the highest value of $v\sin i$-. Filled circles represent the stars in our sample with later spectral types. The theoretical lower limit estimates for each object are derived from Eq. \ref{Eq.W}, and are plotted as open triangles (see text).}
\label{Figure:W10_age_vsini}
\end{figure}

Figure \ref{Figure:W10_age_vsini} relates $<$W$_{10}$(H$\alpha$)$>$ and $v\sin i$ for the stars in the sample \citep[see also][]{Jayawardhana06}. Although there is a scatter at low velocities, all the rapid HAe rotators ($v\sin i$ $>$ 150 km s$^{-1}$) have large H$\alpha$ widths ($<$W$_{10}$$>$ $>$ 600 km s$^{-1}$). Our spectroscopic measurements refer to the non-photospheric contribution (see Paper I). Therefore the maximum H$\alpha$ width must be influenced by the circumstellar gas rotating at the maximum speed, which, for Keplerian rotation, is located in the inner disk. The trend in Fig. \ref{Figure:W10_age_vsini} suggests that the inner disk gas traced by H$\alpha$ is coupled to the rotating stellar photosphere. Magnetospheric channels could be the origin of this link.

\begin{figure}
\centering
 \includegraphics[width=90mm,clip=true]{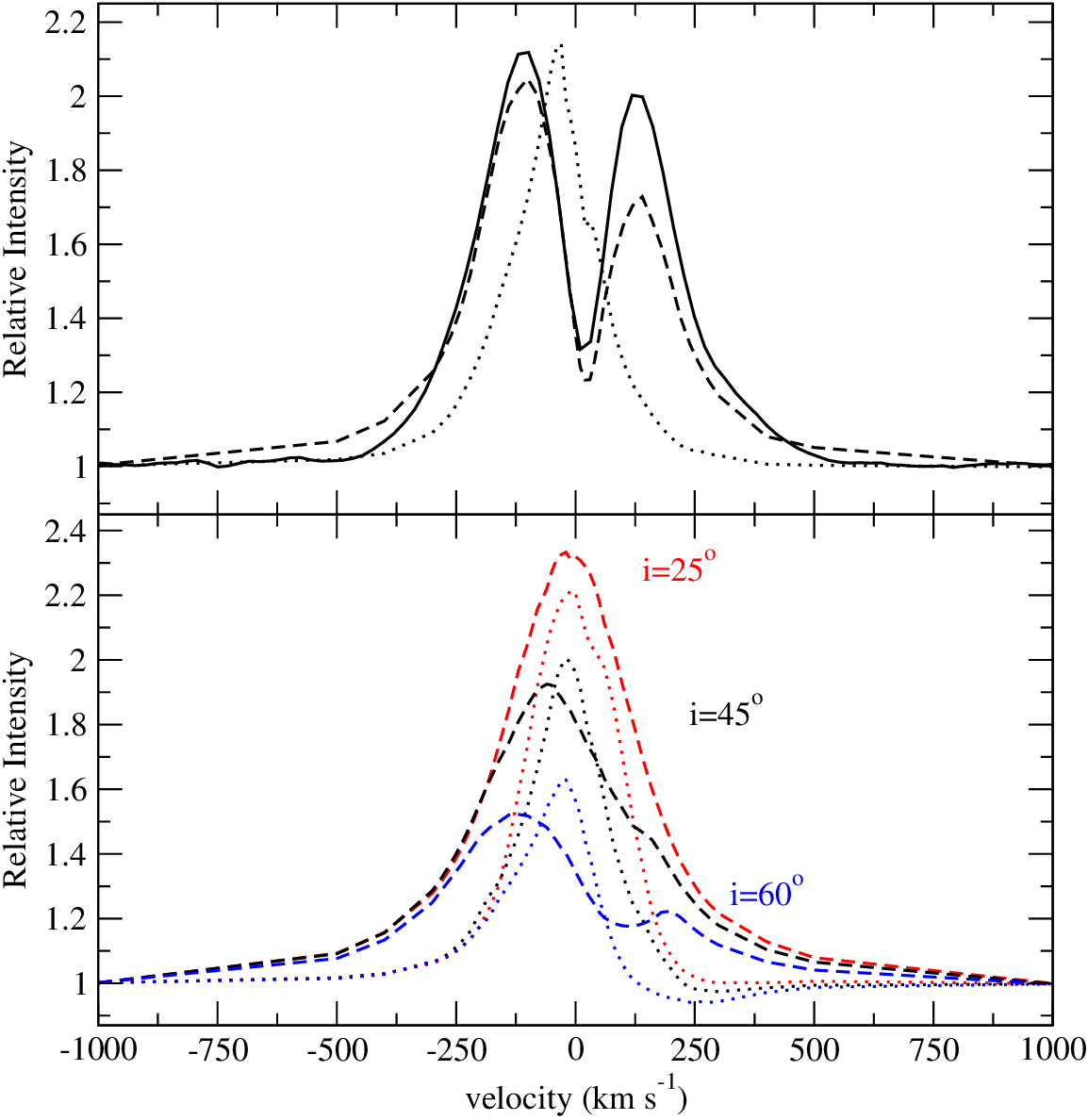}
\caption{(Top): Mean H$\alpha$ profile of \object{BF Ori} from the October campaign in Paper I (solid line) and magnetospheric accretion synthetic profile that best reproduces it (dashed line; M$_*$ = 2.5 M$_{\sun}$, R$_{*}$ = 2.5 R$_{\sun}$, T$_{*}$ = 9000 K, magnetosphere size 4.2--5R$_{*}$, $\rm \dot{M}_{\rm acc}$ = 10$^{-8}$ M$_{\sun}$ yr$^{-1}$, $i$ = 75 deg, and $v\sin i$ = 37 km s$^{-1}$). The dotted line is the same synthetic profile with no rotation. (Bottom): magnetospheric accretion synthetic H$\alpha$ profiles with a stellar rotation rate of 75 km s$^{-1}$ (dashed lines) and with no rotation (dotted lines).} 
\label{Figure:mamrot}
\end{figure}

The influence of the stellar rotation on the H$\alpha$ line broadening can be qualitatively modelled from MA. The MA model described in \citet{Muzerolle01} was successfully used to reproduce several spectral lines of accreting stars with different masses \citep[MDCH04;][]{Calvet04,Muzerolle05}. That assumes a simple dipole geometry, where the H$\alpha$ line is generated from free-falling gas that follows the magnetic lines connecting the magnetosphere in the inner disk and the stellar surface. We applied the model from \citet{Muzerolle01} with input parameters of a typical HAe star. Apart from other effects described in that paper, the H$\alpha$ line broadens whenever the stellar rotation is increased. The strength of the line broadening also depends on additional parameters such as the size of the magnetosphere, the gas temperature, the mass accretion rate, and the inclination to the line of sight. The top panel of Fig. \ref{Figure:mamrot} shows the synthetic profile best fitting the mean H$\alpha$ line shown by \object{BF Ori}. This fit adds to that for \object{UX Ori} in MDCH04, again exemplifying that MA modelling is able to reproduce the line profiles of HAe stars with reasonable accuracy. The modelled profile for \object{BF Ori} includes the corresponding rotation rate \citep[$v\sin i$ = 37 km s$^{-1}$,][]{Mora01}, and is broadened with respect to the same synthetic profile that does not include rotation. The bottom panel of Fig. \ref{Figure:mamrot} plots some other examples of MA synthetic H$\alpha$ profiles showing line broadening as the stellar rotation increases. All profiles were modelled from the same parameters as in the top panel, a magnetosphere size of 2.8--3 R$_{*}$ and different inclinations to the line of sight, as indicated (closer to pole-on for the strongest lines and closer to edge-on for the weakest).

In addition, the trend in Fig. \ref{Figure:W10_age_vsini} is quantitatively consistent with MA operating in HAe stars. When assuming this scenario, the region where the gas is channeled through the magnetic accretion flows has to be located between the stellar photosphere and the co-rotation radius \citep[e.g.][]{Shu94}. At this point the stellar and the disk angular velocities (assuming Keplerian rotation) are equal, deriving:

\begin{equation}
{\rm R}_{\rm cor} \sim \frac{v_{\rm gas}{\rm R}_{*}}{v_{*}} 
\end{equation}
with $v_{gas}$ the azimuthal velocity of the gas at the co-rotation radius. Since $v$$_{*}$ $\geq$ $v\sin i$, and assuming that the non-photospheric H$\alpha$ line width reflects the velocity of the gas at some point between R$_{*}$ and R$_{\rm cor}$, in the sense that $v$$_{\rm gas}$ $\leq$ $<$W$_{10}$(H$\alpha$)$>$/2, we obtain

\begin{equation}
\label{Eq.Rcor1}
{\rm R}_{\rm cor} \lesssim \frac{<{\rm W}_{10}({\rm H}\alpha)>{\rm R}_{*}}{2 v \sin i}.
\end{equation} 

For the $<$W$_{10}$(H$\alpha$)$>$ and $v\sin i$ values plotted in Fig. \ref{Figure:W10_age_vsini}, R$_{\rm cor}$ of HAeBe stars is typically in the range $\sim$ 1--5 R$_*$, which agrees with the values expected for these objects (see Eq. \ref{Eq:Rcor}). We note that R$_{\rm cor}$ $\leq$ 2 R$_*$ for all objects with $v\sin i$ $>$ 150 km s$^{-1}$. Simple dipole geometries would need modifications for such small magnetospheric sizes, in order to reproduce the large line strengths of HAe stars from MA (see MDCH04).

Using Eqs. \ref{Eq:Rcor} and \ref{Eq.Rcor1}, a lower limit for $<$W$_{10}$(H$\alpha$)$>$ can be estimated as a function of the stellar parameters:

\begin{equation}
\label{Eq.W}
<{\rm W}_{10}({\rm H}\alpha)>_{min} \sim 2\left(\frac{GM_*v \sin i}{R_*}\right)^{1/3}.
\end{equation} 
We note that this expression is derived by assuming that the line broadening is determined by Keplerian gas located below the co-rotation radius, which is a major requirement in the MA scenario. 

Figure \ref{Figure:W10_age_vsini} overplots our $<$W$_{10}$(H$\alpha$)$>$$_{min}$ estimates from Eq. \ref{Eq.W}. They reproduce the overall trend followed by our $<$W$_{10}$(H$\alpha$)$>$ measurements. The largest differences between the observed and the estimated values occur for the objects with the lowest values of $v\sin i$. This can be partly attributed to inclination effects, i.e., for pole-on sources $v_{*}$ $>>$ $v\sin i$, and those are expected to be located mainly in the region $v\sin i$ $<$ 150 km s$^{-1}$.

The stars \object{51 Oph}, \object{VV Ser}, \object{VX Cas}, and \object{SV Cep} show $<$W$_{10}$(H$\alpha$)$>$ measurements that are lower than their corresponding $<$W$_{10}$(H$\alpha$)$>$$_{min}$ estimates. The reason could be uncertainties in the stellar parameters, introduced e.g. by close (unknown) companions or simply by the absence of Keplerian-rotating gas between the stellar surface and the co-rotation radius. In this case MA would hardly operate on these objects. Interestingly, all of them are again HBe stars according to their effective temperature ($>$ 10000 $K$). Besides, the B-type object \object{51 Oph} clearly deviates from the general trend in Fig. \ref{Figure:W10_age_vsini}, likely pointing to an inner disk gas decoupled from the central star. The nature of this object is not clear, since it could be a more evolved star or a $\beta$-Pictoris analogue surrounded by a gas-rich debris disk \citep[see e.g.][and references therein]{vandenAncker01,Stark09}. 

In summary, we argue that the high rotational velocities of HAe stars mean that $<$W$_{10}$(H$\alpha$)$>$ cannot be used as tracer of accretion for those objects. At the same time, that the H$\alpha$ line width reflects the stellar rotation can be interpreted as a consequence of MA operating in the late-type HAeBe sources.

\section{Variability of accretion rates and accretion tracers}
\label{section:accretionvar}
The multi-epoch photometry and simultaneous spectra on which this work is based lead us to analyse the accretion rate variability and its possible relation with the changes in the accretion tracers. The photometry and spectra were taken during four campaigns of several days each, on different months \citep[see][and Paper I]{Oudmaijer01}. The multi-epoch values for the Balmer excess were derived from the individual $U-B$ and $B-V$ data taken on the same nights, using an expression analogous to that for $<$$\Delta$D$_B$$>$  in Sect. \ref{section:descriptmodel}. The multi-epoch H$\alpha$ and [\ion{O}{i}]6300 luminosites were derived in paper I from line equivalent widths and the photometric measurements in the $R$ band taken on the same nights. The line luminosities were dereddened using simultaneous values for $B-V$ and a procedure analogous to the one explained in Sect. \ref{Section:Sample} for the mean line luminosities.

Most stars in our sample tend to show a constant Balmer excess, within the uncertainties. The strongest variations are between 0.1 and 0.2 magnitudes, which are detected in some observations of $\sim$ 41$\%$ of the stars. An upper limit for the accretion rate variability can be estimated from the stars showing the smallest mean Balmer excesses and the largest Balmer excess variations (see the right panel of Fig. \ref{Figure:balmeraccretion}). As extreme examples,  \object{V1686 Cyg} shows an increase in the Balmer excess from 0.04 magnitudes on 1998 May 15 to 0.18 magnitudes on 1998 October 27. That change implies an accretion rate increase of a factor $\leq$ 5. The Balmer excess of \object{WW Vul} decreased from 0.14 magnitudes on 1998 May 15 to 0.04 magnitudes on 1998 October 24, which translates into an accretion rate decrease of a factor $\leq$ 4. Considering all the stars, our model yields a typical upper limit of $\sim$ 0.5 dex for the accretion rate changes, on timescales of days to months. More accurate constraints on possible accretion rate variations below our detection limit would require UV spectra or narrow-band photometry.

\begin{figure}
\centering
 \includegraphics[width=90mm,clip=true]{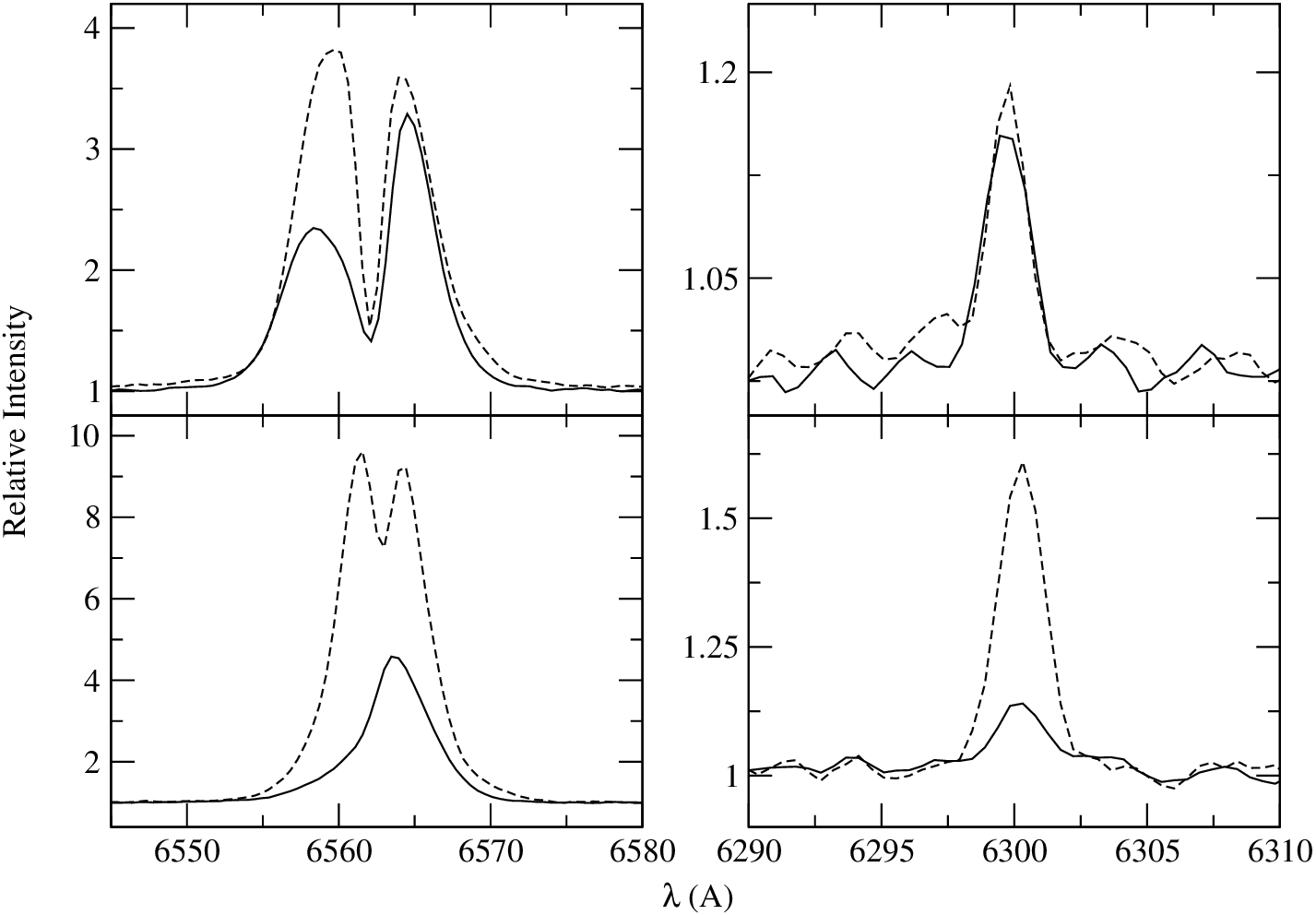}
\caption{H$\alpha$ (left panels) and [\ion{O}{i}]6300 (right panels) normalized profiles for \object{RR Tau} (top) and \object{HK Ori} (bottom), with different scales depending on the panel. The dashed and solid lines represent the first and second observing dates (see text).} 
\label{Figure:Balmer_lines_simul}
\end{figure}

The analysis of the simultaneous variability of the Balmer excesses and the line luminosities is not only limited by the error bars provided for our broad-band photometry and mid-resolution spectra, but also by the number of data sampled per object. There is a maximum of 7 simultaneous photometric and spectral points per star, being typically 3 (see Table 3 in Paper I). However, we find objects showing changes in the Balmer excess for which the corresponding H$\alpha$ and [\ion{O}{i}]6300 luminosities do not vary accordingly, as would be expected for an accretion tracer. As an example, \object{V1686 Cyg} shows a Balmer excess close to 0 magnitudes on 1998 May 15, which increased to the typical value from our averaged data (0.12 magnitudes) the following night. However, the corresponding H$\alpha$ and [\ion{O}{i}]6300 luminosities did not change. In contrast, there are stars showing variations in the line luminosities with a constant value of the Balmer excess. The one shown by \object{BH Cep} was $\sim$ 0 magnitudes during all observations, but the H$\alpha$ and [\ion{O}{i}]6300 luminosities changed by a factor of almost 2. Two other stars showing a decoupled behaviour between the Balmer excess and the line luminosities are provided in Fig. \ref{Figure:Balmer_lines_simul}. \object{RR Tau} increased the Balmer excess from $\sim$ 0.32 magnitudes on 1998 October 27, to 0.45 magnitudes on 1999, January 30. Simultaneously, the H$\alpha$ luminosity decreased by a factor 1.4 and the [\ion{O}{i}]6300 luminosity remained constant within the uncertainties -note that the slight variation in the equivalent width was not accompanied by a change in the line luminosity (Paper I)-. \object{HK Ori} showed a constant Balmer excess of $\sim$ 0.57 magnitudes at the same time that the H$\alpha$ and [\ion{O}{i}]6300 luminosities decreased by a factor $\sim$ 2 from 1998 October 25 to the following night. The examples in this sense are not scarce, but the inverse case is not true; we do not detect stars in our sample for which a significant increase or decrease in the excess is reflected by the corresponding one in the line luminosities, but again, we note that most possible variations fall below our detection limit.

Even considering the limitations of our data to analyse such small changes, the results described above suggest that the H$\alpha$ and [\ion{O}{i}]6300 luminosities do not trace the accretion rate variability in several stars, unless they reflect the accretion rate changes with a certain lag not covered by our data. Night-to-night variations in the atmospheric transmission could contaminate the estimates of the variability in the Balmer excess. However, a careful photometric calibration was made, especially in the $U$ band, by observing standard stars with different spectral types and airmasses and making several tests after the calibrations \citep{Oudmaijer01}. It is unknown whether the variations in other empirical tracers not analysed in this work could correlate with the accretion rate variability from the changes in the Balmer excess, or whether the variability could be affected by other factors such as stellar companions. These topics are the subject of ongoing investigation. Multi-epoch, high-resolution spectra covering simultaneously the wavelength range from the UV to the near-IR will be very valuable in this sense. Those can be taken with the X-Shooter spectrograph on the VLT. 

\section{Discussion}
\label{Section:Discussion} 

The origin of the empirical calibrations between accretion and line luminosities could be explained from the influence that accretion has on the line formation: Both the H$\alpha$ and Br$\gamma$ lines have been explained from MA models \citep[][MDCH04]{Muzerolle98b,Muzerolle01}, although the influence of winds can be non-negligible \citep{Kurosawa06,Kraus08}. The latter contribution seems to be more important as the stellar mass increases (see Paper I and references therein), which would explain the possible decrease in the slope of the H$\alpha$ calibration, when compared to that for lower mass stars (left panel in Fig. \ref{Figure:emcor}). Accretion-powered outflows could be the origin of the [\ion{O}{i}]6300 feature \citep{Corcoran97,Corcoran98}. They could therefore link the luminosity of this line with that of accretion (middle panel in Fig. \ref{Figure:emcor}). If the [\ion{O}{i}]6300 emission is caused by the UV flux illuminating the disk surface \citep{Acke05}, then the UV excess from the accretion shock would explain the trend with the [\ion{O}{i}]6300 luminosity. However, several HAe stars with strong Balmer excesses do not show [\ion{O}{i}]6300 emission, regardless of their different degrees of disk flaring according to the \citet{Meeus01} groups (see e.g. the group II object \object{HD 150193} and the group I star \object{V346 Ori}).

Although the mean H$\alpha$ and [\ion{O}{i}]6300 luminosities are related to the ``typical'' accretion rates (Sect. \ref{section:actracers}), the numerous cases for which the changes in the line luminosities and in the simultaneous Balmer excesses are decoupled suggest that those lines do not to trace the accretion rate variability. This could be the case for several other lines because the typically low accretion rate changes we have derived -less than 0.5 dex- contrast with the complex variability behaviour shown by the spectral tracers. Multi-epoch observations indicate that their variability amplitudes are not correlated: the \ion{He}{i}5876 line shows equivalent width variations over one order of magnitude for a given HAe star, at the same time that H$\alpha$ varies typically by less than a factor 4 (Paper I). \citet{Nguyen09} reports that the \ion{Ca}{ii} fluxes provide typical accretion rate variations of 0.35 dex while the H$\alpha$ 10$\%$ width suggests changes of 0.65 dex for the same sample of CTT stars. These differences could be extended to a wide number of empirical tracers of accretion covering a wide wavelength range, indicating that the origin and strength of several of these lines are most probably influenced by different processes apart from accretion.

These results make it reasonable to doubt that several empirical calibrations between the ``typical'' accretion and line luminosities could not be caused by the accretion contribution to the lines but by a different source. We tentatively suggest that a common dependence of both the accretion and the line luminosities on the stellar one could have strong influence, thereby driving the empirical calibrations. First, there seems to be a direct dependence of $L$$_{\rm acc}$ on L$_{*}$ (Fig. \ref{Figure:miaccretion}, right). Second, it seems from the comparison between the data for low-mass stars and our results \citep[see e.g.][and Fig. \ref{Figure:emcor}]{Herczeg08} that several line luminosities are also correlated with L$_{*}$; e.g., most low-mass stars show H$\alpha$ luminosities in the range $-6 \leq \log ({L}_{{H}\alpha}/{L}_{\sun})
\leq -2$, orders of magnitude lower than those shown by our more luminous stars. In addition, the line luminosities of the objects for which the shock models are not able to reproduce the Balmer excesses (see Sect. \ref{section:discuss}) tend to be the highest in our sample, as are their stellar luminosities (see Table \ref{Table:sample}). From a statistical point of view, the Spearman's probability \citep[see e.g.][]{Conover80} that our accretion and H$\alpha$ luminosities are uncorrelated in Fig. \ref{Figure:emcor} is only p = 4.1$\times$10$^{-4}$$\%$. However, this value does not consider the possible common dependence of both luminosities on L$_{*}$. The partial correlations technique \citep[e.g.][]{Wall03} is useful for this task, since it measures the degree of correlation between two variables, nullifying the effect of a common dependence on a third one. Under the hypothesis that both $L$$_{\rm acc}$ and L(H$\alpha$) depend on L$_{*}$, the Spearman's probability from partial correlations increases to 27$\%$. This value
is close to that obtained from the comparison of
two sets of random variables. A similar conclusion is reached for the Br$\gamma$ calibration and, to a lesser extent, for the [\ion{O}{i}]6300 one. These results indicate that the $L$$_{\rm acc}$ -- L$_{line}$ relations become much less significant once the common dependence on L$_{*}$ is removed. In this view, the apparent decrease in the slopes in the  $L$$_{\rm acc}$ -- L$_{line}$ trends for the intermediate-mass regime (Fig. \ref{Figure:emcor}) would be reflecting the possible decrease in the $L$$_{\rm acc}$ -- L$_{*}$ correlation (right panel in Fig. \ref{Figure:miaccretion}). 

\section{Conclusions}
\label{Section:Conclusions}
We applied shock modelling within the context of MA to 38 HAeBe stars, reproducing the strength of the observed Balmer excesses and deriving accretion rate estimates. The typical mass accretion rate is 2 $\times$ 10$^{-7}$ M$_{\sun}$ yr$^{-1}$, with scatter and a steep dependence on the stellar mass. The latter can be explained by the most massive HAeBe stars being the youngest, hence the strongest accretors. We obtained empirical expressions to relate accretion and the H$\alpha$, [\ion{O}{i}]6300, and Br$\gamma$ luminosities, which are similar to those for the lower mass regime. However, there could be a slight decrease in the slope of the trends relating the accretion luminosity with both the line and the stellar luminosities. In contrast, the H$\alpha$ line width at 10$\%$ of peak intensity can usually not be used to estimate accretion rates for HAeBe stars, unlike in the lower mass regime. The H$\alpha$ width of HAe objects broadens as the projected rotational velocities increase, which is also consistent with MA operating in these stars.

Although the ``typical'' accretion rates show clear trends with the mean line luminosities, the accretion-rate changes from the Balmer excess -typically lower than 0.5 dex- seem to be uncorrelated with the variability of the H$\alpha$ and [\ion{O}{i}]6300 lines. We tentatively suggested that the origin of the empirical calibrations between the accretion and line luminosities could not be driven by the influence of accretion on the emission lines, but by a common dependence on the stellar luminosity. The statistical level of significance of the empirical calibrations is strongly diminished once this dependence is considered.

The shock model used is not able to reproduce the strong Balmer excesses shown by the four hottest stars in our sample under reasonable input parameters. In addition, the H$\alpha$ 10$\%$ width of four HBe objects is not consistent with the presence of Keplerian gas between the stellar surface and the co-rotation radius, which is a major requirement for MA to operate. These results give additional support for a change from magnetically driven accretion in HAe stars to some other kind of accretion in HBes \citep[see e.g.][]{Vink02,Eisner04,Mottram07}. 

It is finally pointed out that, although our results were obtained and interpreted within the context of MA, this means neither that the authors found a way to test magnetospheric accretion onto HAeBe stars nor that the observational data could not be explained from a different scenario. In this respect, further observational and theoretical efforts on the magnetic fields necessary to drive MA in these objects are needed to establish firmer conclusions.

\begin{acknowledgements}
The authors thank the referee, Greg Herczeg, for his useful comments on the original manuscript, which helped us to improve
the paper. C. Eiroa, I. Mendigut\'{\i}a, and B. Montesinos are partially supported by grant AYA-2008 01727. I. Mendigut\'{\i}a is grateful for financial support from the Space Telescope Science Institute during a two-month stay. This research made use of the SIMBAD database, operated at the CDS, Strasbourg, France 
\end{acknowledgements}

\appendix

\end{document}